\documentclass[aps,pre,showpacs,twocolumn,eqsecnum,floatfix,tighten,superscriptaddress]{revtex4-1}
\usepackage{amsmath}
\usepackage{amsthm}
\usepackage{amssymb}
\usepackage{amsbsy}
\usepackage{graphicx}
\usepackage{dcolumn} 
\usepackage{bm, bbm} 
\usepackage{txfonts}
\usepackage{epic}
\usepackage{wasysym}
\usepackage{epsf, times}
\usepackage{color}
\usepackage[breaklinks=true,urlcolor=blue,bookmarks=true]{hyperref}

\usepackage{esint}

\definecolor{darkblue}{rgb}{.0, .0,.7}
\hypersetup{pageanchor=false}

\begin{document}
\title{Low-frequency dielectric response of a periodic array of charged spheres in an electrolyte solution: The simple cubic lattice}
\author{Chang-Yu Hou}
\affiliation{Schlumberger-Doll Research, 1 Hampshire Street, Cambridge, MA 02139, USA}
\author{Jiang Qian}
\affiliation{4097 Silsby Road, University Heights, OH 44118, USA}
\author{Denise E. Freed}
\affiliation{Schlumberger-Doll Research, 1 Hampshire Street, Cambridge, MA 02139, USA}

\date{\today}

\begin{abstract}
We study the low-frequency dielectric response of highly charged spheres arranged in a cubic lattice and immersed in an electrolyte solution. We focus on the influence of the out-of-phase current in the regime where the ionic charge is neutral. We consider the case where the charged spheres have no surface conductance and no frequency-dependent surface capacitance. Hence, the frequency dispersion of the dielectric constant is dominated by the effect of neutral currents outside the electric double layer. In the thin double-layer limit, we use Fixman's boundary condition at the outer surface of the double layer to capture interaction between the electric field and the flow of the ions. For periodic conditions, we combine the methods developed by Lord Rayleigh for understanding the electric conduction across rectangularly arranged obstacles and by Korringa, Kohn and Rostoker for the electronic band structure computation. When the charged spheres occupy a \emph{very} small volume fraction, smaller than one percent, our solution becomes consistent with the Maxwell Garnett mixing formula together with the single-particle polarization response, as expected, because inter-particle interactions become less prominent in the dilute limit.  By contrast, the inter-particle interaction greatly alters the dielectric response even when charged spheres occupy only two percent of the volume. We found that the characteristic frequency shifts to a higher value compared to that derived from the single-particle polarization response. At the same time, the low-frequency dielectric enhancement, a signature of charged spheres immersed in an electrolyte, becomes less prominent for the periodic array of charged spheres. Our results imply that the signature of the dielectric response of a system consisting of densely packed charged spheres immersed in an electrolyte can differ drastically from a dilute suspension.
\end{abstract}

\maketitle

\section{Introduction}
\label{introduction}

Immersing charged particles and flakes in an electrolyte often results in an interesting electromagnetic response. For instance, a strong low-frequency dielectric enhancement accompanied by a non-Lorentzian relaxation frequency dependence is observed experimentally~\cite{Schwan1962,Chassagne2003}. It had been argued by Schwan et. al.~\cite{Schwan1962} that the conventional Maxwell-Wagner relaxation, frequency-independent surface conductance, and electrophoresis cannot fit their experiments results. The dielectric enhancement strength depends not only on the amount of charges but also on the size and shape of particles~\cite{Schwarz1962,Dukhin1974,Fixman,Dukhin1980,DeLacey1981,Chew1982,Chassagne2008,Chassagne2013,Hou2017,Hou2018}. The frequency dispersion of its dielectric response also offers a scheme to probe the microscopic state of suspended colloids and particles in the solution~\cite{Schwan1962,LEROY2017,Hou2018}. In addition, some electronic devices, such as super-capacitors, are constructed from the similar physical system, where charged particles are densely packed into the electrolyte. Hence, better theoretical understandings of the dielectric response of such system will pave ways for new technologies and applications.

The theoretical study of the low-frequency dielectric response of charged 
particles immersed in an electrolyte often start from obtaining the solution 
of the single particle polarization coefficient~\cite{Sihvola1999}. Because 
ions are responsible for the conduction in electrolytes and form the so-called 
electric double-layer structure around a charged particle, solving for the 
polarization coefficient requires the consideration of diffusive ionic flows in 
the electrolyte, consistent with the electric potential. Hence, compared to a 
similar classical electrodynamics problem where a dielectric ball is embedded 
into another dielectric material~\cite{Jackson1998,Landau1984}, the solution of 
the polarization coefficient for a charged sphere immersed in an electrolyte 
becomes nontrivial~\cite{Dukhin1974,Fixman,Dukhin1980,Chew1982}. Based on 
different single-particle polarization responses, various physical mechanisms 
have been proposed for qualitatively explaining the observed dielectric dispersion 
and enhancement~\cite{Schwarz1962,Chew1982,Chassagne2003}.

Two main mechanisms are often used for understanding the experimentally observed phenomena. First, a frequency-dependent surface complex conductance, based on the diffusion of counter-ions constrained on the surface, was used to explain the unusual dielectric enhancement~\cite{Schwarz1962}. On the other hand, the out-of-phase neutral ionic current, outside the double-layer, can be another mechanism leading to the strong low-frequency dielectric enhancement~\cite{Chew1982,Chew1982a}. Both mechanisms predict a characteristic relaxation frequency associated with the particle size. The former mechanism requires the introduction of particle size or activation energy distributions to construct a broader non-Lorentzian frequency response for the relaxation. In contrast, the latter naturally gives rise to the non-Lorentzian relaxation as observed in experiments. 

With the single-particle polarization coefficient, some types of effective 
medium approximation, such as Maxwell Garnett~\cite{Garnett1904,Garnett1906} 
and Bruggeman~\cite{Bruggeman1935} mixing formulas, are often used for modeling 
the eventual dielectric response of a mixture. In general, these effective 
medium theories aim to take into account the effect of surrounding dipoles for 
the included materials and yield different degrees of successes for mixtures of 
dielectric materials~\cite{Sihvola1999,Milton2002}. Because the diffusive ionic 
flows far from the electric double-layer can dominate and lead to a strong 
low-frequency dielectric enhancement~\cite{Chew1982}, diffusive ionic flows 
centered around a charged particle will be affected by the presence of other 
charged particles, an effect which cannot be captured by effective medium 
approximations.  As a result, it is of interest to study, even in an idealized 
setup, the dielectric response of charged particles immersed in electrolyte 
beyond the usual effective medium approximation usied with the single particle 
polarization response.

In this paper, we will investigate the low-frequency dielectric response of charged spheres forming a simple cubic lattice immersed in the electrolyte. Our study focuses on the dielectric enhancement mechanism due to the presence of the out-of-phase current proposed by Chew and Sen~\cite{Chew1982}. The static limit of a similar construction has been studied by Sen and Kan for a two-dimensional array~\cite{SenKan1,SenKan2} and by Sen for the simple cubic lattice~\cite{Sen1987}. Because a strict zero-frequency limit was taken, only values of the real static conductivity are obtained in the prior studies. As a result, the frequency dependence of the complex dielectric constant (or equivalently conductivity) and the low-frequency dielectric enhancement for a periodic array of charged spheres immersed in the electrolyte are still open questions.

Because our focus in this work is on how the neutral ionic currents impact the dielectric response, our model will assume that there is no surface conduction on the charged spheres. Investigations into the effect of the surface conduction on the dielectric response of the periodic structures can be found in Refs.~\cite{Gu1991} and~\cite{Ni2002}. In those papers, the dielectric response of a periodic array of charged spheres does not significantly deviate from that obtained from sthe effective medium model with the single-particle polarization response because the mechanism leading to the dielectric enhancement there comes from the surface conduction. 

In order to capture the effect of the neutral ionic currents, we follow Fixman's 
thin double-layer approximation for deriving a pair of effective boundary 
conditions just outside the electric double-layer~\cite{Fixman}. These boundary 
conditions couple the perturbed potential and ion concentrations, due to the 
driving external field. We then use Lord Rayleigh's method, later corrected 
and expanded upon by McPhedran and Mckenzie, for deriving the periodic condition for 
the perturbed potential~\cite{Rayleigh1892,McPhedran1,McPhedran2}. In addition, 
the Korringa, Kohn and Rostoker (KKR) method, developed for band structure 
calculations in solids, is used to derive a periodic condition, with a 
vanishing Bloch vector, for the perturbed ion
concentration~\cite{Korringa,KohnRostoker}.  Combining these three parts gives 
a set of coupled linear equations that provide the \emph{complete} functional 
form of the perturbed potential and ion concentrations, up to arbitrary order 
of the angular momentum number $\langle l,m\rangle$, in the electrolyte 
solution outside the double layer. Finally, Lord Rayleigh's Green's theorem 
trick~\cite{Rayleigh1892} is used to extract the complex conductivity from the 
$l=1,m=0$ coefficient in the expansion of the perturbed potential 
solution~\cite{Rayleigh1892}.

The paper is organized as follows. In Section~\ref{sec:model}, we discuss the fundamental governing equations and Fixman's thin double layer approximation for the electric potential and the ion concentration. In Section~\ref{sec:Periodic_BCs}, periodic boundary conditions associated with the perturbed electric potential, using Lord Rayleigh's method, and with the ion concentrations, using the KKR method, are developed. In Sec.~\ref{sec:conductivity_B10}, we first establish the link between the expansion coefficients of the electric potential in terms of spherical harmonics to the complex conductivity of the model. We then derive the formula for the complex conductivity to the lowest order in the angular momentum cutoff. In Sec.~\ref{sec:conductivity_higher_order}, we present the complex conductivity evaluated with higher order angular momentum cutoffs. We illustrate the scheme for setting up numerical evaluations and demonstrate the convergent result of the complex conductivity to numerically accessible orders. Qualitatively, our studies show that the single-particle polarization response cannot be used to represent the dielectric response of the system even with a fairly small volume fraction,~$2$\%, occupied by the charged spheres. We provide some discussion on the implications of our results in Sec.~\ref{sec:discussion} and then a brief summary in Sec.~\ref{sec:summary}. In Appendix~\ref{app:structure-factor}, we review the scheme for evaluating the "structure factors", geometrical constants required for complete the periodic boundary conditions in our computation.

\section{Formulation of the Model and Boundary Conditions}
\label{sec:model}

We consider a system consisting of identical charged spheres with radius $a$ arranged in a simple cubic lattice with spacing $b$ immersed in an electrolyte solution. The surface charge of the spheres is uniformly distributed around each sphere. An electric double layer dominated by the counterions with thickness of a few Debye lengths, $\lambda=\sqrt{\epsilon_w k_B T/(2 e_0^2 N_0)}$, will surround each charged sphere. Here, $\epsilon_w$ is the permittivity of the electrolyte, $N_0$ is the equilibrium ion concentration in the electrolyte, $e_{0}$ is the absolute value of the electron charge, $k_B$ is the Boltzmann constant, and $T$ is the temperature of the electrolyte. For simplicity, we will consider monovalent ions in this paper. Considering that $\lambda$ is usually in the range of a nanometer, the thin double-layer condition $\lambda \ll a$ is easily satisfied even for sub-micron sized particles. The surface charge of the spheres is fully screened by the net charge in the double-layer. Hence, the electrolyte remains charge neutral outside the double-layer.

The hydrodynamic flows are assumed to be small compared to the conductive currents.  This is a valid assumption if the net potential drop across the double layer is comparable to or smaller than $k_B T/e_0$~\cite{jiang}. Furthermore, Fixman has shown that electrophoresis flows only add qualitative modifications to the polarization response due to ionic current flows~\cite{Fixman1983}. Hence, we will only focus on the direct ionic currents that consist of the conductive and the diffusive currents. Both the cation and anion currents, $J_{\pm}$, separately obey continuity equations given by
\begin{equation}
\label{eq:total-current}
\begin{split}
	\overrightarrow{\nabla}\cdot\vec{J_\pm} &=-\frac{\partial 
	N_\pm}{\partial t}, \quad	\mu_\pm=\ln N_\pm\pm\frac{e_0\Psi}{k_B T},
\\
	\vec{J}_{\pm}&=-D N_{\pm}\overrightarrow{\nabla}\mu_\pm
	=-D\left(\overrightarrow{\nabla}N_{\pm}\pm\
	\frac{e_0 N_{\pm}}{k_B T}\overrightarrow{\nabla}\Psi\right).
\end{split}
\end{equation}
Here, $N_\pm$ are the ionic densities for the cations and anions, $\Psi$ is the potential, and $\mu_\pm$ are the ionic chemical potentials in the electrolyte. For simplicity, the diffusion coefficients of both the cations and anions, $D$, are assumed to be equal.

Now, subject the system to a small and quasi-static~\footnote{The condition is, as always, that the wavelength of the electromagnetic perturbation, which for a MHz drive is hundreds of meters, is much larger than the system's geometric features.} periodic driving electric-field, $\vec{E}=H\exp(-i\omega t)$. We can 
rewrite the ionic densities and the potential as follows:
\begin{equation}
\begin{split}
\label{eq:perturbation}
	N_\pm &=N^{eq}_\pm+n_\pm,\quad\Psi=\Psi^{eq}+\psi,
	\\
	\vec{J}_\pm&\approx-D\left[\overrightarrow{\nabla}n_\pm\pm
	\frac{e_0}{k_B T}
	(N^{eq}_\pm\overrightarrow{\nabla}\psi+ n_\pm \overrightarrow{\nabla} \Psi^{eq})
	\right]
\end{split}	
\end{equation}
Here, the superscript ``$eq$'' denotes equilibrium values in the absence of the driving field $\vec{E}$. It is assumed that the perturbations due to the electric field are small: $n_\pm\ll N^{eq}_\pm$ and $\psi\ll\Psi^{eq}$. Thus, we neglect the second order terms in the equation for the current. We also used the fact that in equilibrium, both currents vanish by definition.

The potential satisfies the Poisson equation that relates it to the charge density.  Because of superposition,
the perturbed potential $\psi$ and perturbed charges $n_+$ and $n_-$ will separately satisfy the Poisson
equation given by
\begin{equation}
\overrightarrow{\nabla}\cdot \epsilon\overrightarrow{\nabla}\psi = -e_0 (n_+-n_-). 
\label{eq:psi_np_nm}
\end{equation}
In this equation, $\epsilon$ is the dielectric constant given by $\epsilon_w$ outside the particle and
$\epsilon_p$ inside the particle.

Outside of the double layer, the equilibrium ion densities simply equal the bulk ion densities $N^{eq}_\pm=N_0$. The equilibrium potential obeys the simple Laplace's equation $\nabla^2\Psi^{eq}=0$. In addition, the electrolyte always remains neutral outside the double layer, with or without the driving field~\cite{Fixman} (A proof showing this condition with our current notation can be found in Ref.~\onlinecite{Hou2017}). As a result, $n_+=n_- \equiv n$. Furthermore, the ionic currents simplify greatly 
outside the double layer:
\begin{equation}\label{eq:current-beyond-DL}
	J_\pm=-D\left(\overrightarrow{\nabla}n\pm
	\frac{e N_0}{k_B T}\overrightarrow{\nabla}\psi
	\right).
\end{equation}
This equation, combined with the ion continuity equation, gives us a remarkably 
simple pair of equations that characterize \emph{all} the physics beyond 
the double-layer:
\begin{equation}\label{eq:governing-beyond-DL}
	\nabla^2\psi=0,\qquad \nabla^2 n= - \frac{i\omega}{D}n.
\end{equation}
Here, we have written the governing equations in the frequency domain, where $\omega=2 \pi \nu$ is the radial frequency with $\nu$ being the regular frequency. 

The strategy forward is then to incorporate \emph{all} the physics \emph{within} the double-layer into a pair of boundary conditions relating $\psi$ and $n$ at the edge of the double-layer $r=a+\delta$. For that, we integrate the continuity equation along the radial spherical 
coordinate centered on one of the spheres~\cite{Fixman}:
\begin{equation}
\label{eq:radial-integral}
	i\omega\int_{a}^{a+\delta} n_\pm~dr =J_{\pm}^r|_{a+\delta}+
	\int_a^{a+\delta}
	\overrightarrow{\nabla}_{\parallel}\cdot J^{\parallel}_\pm~dr.
\end{equation}
$J_{\pm}^r$ and $J^{\parallel}_\pm$ are ionic currents normal (in the radial direction) and parallel to the surface of the sphere, respectively. Here, we have used the boundary condition that the normal ionic current must vanish at the sphere's surface, $r=a$. By using Eq.~\eqref{eq:total-current}, we can further simplify the parallel part of the divergence as 
\begin{equation}
\label{eq:parallel}
\begin{split}
	\overrightarrow{\nabla}_{\parallel}\cdot J^{\parallel}_\pm &=
	-D\left[\overrightarrow{\nabla}_{\parallel} N_\pm\cdot
	\overrightarrow{\nabla}_{\parallel}\mu_\pm+
	N_\pm\overrightarrow{\nabla}_{\parallel}\cdot\overrightarrow{\nabla}_{\parallel}\mu_\pm
	\right]
	\\
	&\approx -D N_\pm\overrightarrow{\nabla}_{\parallel}\cdot\overrightarrow{\nabla}_{\parallel}\mu_\pm.
\end{split}
\end{equation}
To derive the last step, we note that the physics around each sphere remains radially symmetric in the absence of the driving field $\vec{E}$ even in a periodic array, as long as the charged spheres are separated by at least several Debye lengths, or equivalently, the double layers around different spheres do not overlap. This is because, under this assumption, easily satisfied in all but the close-packed geometry, the screening by the double layer around each sphere ensures the electric field from surface charge on that sphere does not directly affect the charge dynamics on other spheres nearby. From such radial symmetry we have, in equilibrium, $\overrightarrow{\nabla}_{\parallel} N^{eq}=0$ and $\overrightarrow{\nabla}_{\parallel} \Psi^{eq}=0$. Because both $\overrightarrow{\nabla}_{\parallel}\mu_\pm\approx(\overrightarrow{\nabla}_{\parallel}
n_\pm)\pm (e/k_B T)\overrightarrow{\nabla}_{\parallel}\psi$ and $\overrightarrow{\nabla}_{\parallel}
N_\pm=\overrightarrow{\nabla}_{\parallel} n_\pm$ are first order in the perturbation, their product is a second order term that can be safely neglected.

Furthermore, we note that $\overrightarrow{\nabla}\mu_\pm$ are conservative vector fields (their curl is zero) and thus their tangential component is uniform over short distances in the radial direction as long as their radial components are finite~\footnote{The proof is standard and involves integrating the vector field around a loop,
	a rectangle that is long in the tangential direction and very short,
	across $\delta$ in the radial direction, and that the short radial line
	integrals can be neglected. However, it may be objected that within
	the double layer there is a huge gradient in the potential because of the
	build-up of charges within the short distance $\delta$, so the radial
	integrals may be non-negligible. However, from Eq.~\eqref{eq:perturbation}
	we can see that the radial gradient of the \emph{chemical potential}
	$\mu$ is actually of the order of the perturbation, due to the absence 
	of currents in equilibrium. Thus the usual argument indeed works, as long
	as the right order of limits is taken.}. This means that we can treat $\overrightarrow{\nabla}_{\parallel}\cdot\overrightarrow{\nabla}_{\parallel}\mu_\pm$ in Eq.~\eqref{eq:parallel} as a constant and take it out of the radial integrals in Eq.~\eqref{eq:radial-integral}. Assuming $D$ is constant across the double layer, these integrals simply give the ion densities per unit area in the double layer. Let us assume that the negative surface-charge density on the sphere, $\Omega$, is so large that the ions in the double layer are predominantly cations~\cite{Bourg2011}. Then it follows from the 
perfect screening of this surface charge by the double layer that
\begin{equation}\label{eq:parallel-integrated}
	\int_a^{a+\delta}
	\overrightarrow{\nabla}_{\parallel}\cdot J^{\parallel}_+~dr
	\approx -D \Omega \overrightarrow{\nabla}_{\parallel}\cdot\overrightarrow{\nabla}_{\parallel}\mu_+,\quad
	\int_a^{a+\delta}
	\overrightarrow{\nabla}_{\parallel}\cdot J^{\parallel}_-~dr
	\approx 0.
\end{equation}
Because $
\overrightarrow{\nabla}_{\parallel}\cdot\overrightarrow{\nabla}_{\parallel}\mu_+
$ is uniform across the double layer, we can choose its value to be evaluated
right at the outer boundary $r=a+\delta$, just like that of $J^{r}_\pm$ in Eq.~\eqref{eq:radial-integral}. When the diffusion coefficient $D$ is not constant, a weighted surface cation density can be used, instead, for the rest of this discussion~\cite{Sen1987}.

Finally, the only term not yet expressed in terms of physical quantities outside the double layer is the left-hand side of Eq.~\eqref{eq:radial-integral}. This term, which represents the radial integral of $n_\pm$, turns out to be negligible. To show this, it suffices to show that it is much smaller than the radial current.  We expect
$n_-$ to be very small inside the double layer, so that, according to Eqs.~\eqref{eq:radial-integral} and \eqref{eq:parallel-integrated}, the radial current for the negative ions at the outer boundary of the double layer should be close to zero.  That means that the radial derivative of the perturbed charge and of $eN_0/k_BT$ times the perturbed potential are similar in size.  Because the radial current for the positive ions is the sum of these two terms, we need only to determine when one of these terms is significantly larger than the left-hand side of Eq.~\eqref{eq:radial-integral}. To do so, we will follow Fixman's argument and integrate Poisson's equation for the perturbed potential, Eq.~\eqref{eq:psi_np_nm}, from just inside the surface of the particle to the outer edge of the double layer to obtain
\begin{equation}
\epsilon_w \left. \frac{\partial \psi}{\partial r}\right|_{r=a+\delta} 
 - \epsilon_p \left. \frac{\partial \psi}{\partial r}\right|_{r=a^-} \approx -e\int_a^{a+\delta} n_+ dr.
\end{equation}
In this equation, we have neglected the integral over the tangential derivative of $\psi$.  When the dielectric
constant inside the particle is zero, we then have 
\begin{equation}
-D\frac{eN_0}{k_BT}\left. \frac{\partial\psi}{\partial r}\right|_{r=a+\delta} \approx \frac{D}{\delta^2}\int_a^{a+\delta} n_+ dr,
\end{equation}
where we have used the fact that the double layer thickness $\delta$ is in the same order of the Debye length $\lambda$. Thus, the left-hand side of Eq.~\eqref{eq:radial-integral} can be ignored when $\omega << D/\delta^2$. When the dielectric constant inside the particle, $\epsilon_p$ is not zero, this bound will decrease.

By combining Eqs.~\eqref{eq:radial-integral} and~\eqref{eq:parallel-integrated}, and neglecting the smaller terms at low 
frequency, we have the following effective boundary conditions at the outer 
edge of the double layer:
\begin{equation}\label{eq:BC-outside-DL}
	J^r_-|_{r=a+\delta}=0,
	\quad
	J^r_+|_{r=a+\delta}-D\Omega~
	\overrightarrow{\nabla}_{\parallel}\cdot\overrightarrow{\nabla}_{\parallel}\mu_+\big|_{r=a+\delta}
=0.
\end{equation}
Now, using Eqs.~\eqref{eq:total-current} and~\eqref{eq:current-beyond-DL}, we can explicitly write the two equations as:
\begin{equation}
\label{eq:BC-explicit}
\begin{split}
	0 & = \left. \frac{\partial n}{\partial r}-\frac{e N_0}{k_B T}
	\frac{\partial \psi}{\partial r}\right|_{r=a+\delta},
	\\
    0 & =\left. -\left(\frac{\partial n}{\partial r}+\frac{e N_0}{k_B 
	T}\frac{\partial \psi}{\partial r}\right)-\frac{\Omega}{N_0}\frac{L^2}{r^2}\left(
n+\frac{e N_0}{k_B T}\psi
\right)\right|_{r=a+\delta}.
\end{split}
\end{equation}
Here, we introduced the usual shorthanded $L$ for the angular momentum operator that is the tangential component of the Laplace operator: $L^2/r^2=\nabla^2-\partial^2/\partial r^2$. These two boundary conditions at $r=a+\delta\approx a$, combined with the equations of motion Eqs.~\eqref{eq:governing-beyond-DL} for the ion density $n$ and the potential $\psi$, completely characterize the physics beyond the double layer. In the following section, we will discuss the periodic conditions required for solving these equations when the charged spheres are arranged in a simple cubic lattice. 

Let us set the origin of the spherical coordinate system at the center of 
one sphere.  Then the general solutions for $n$ and $\psi$ in Eqs.~\eqref{eq:governing-beyond-DL} are:
\begin{eqnarray}
\label{eq:general-solutions_psi}
\psi&=&\sum_{l=0}^{\infty}\sum_{m=0}^{l}\left[
A_{lm} r^l + B_{lm} r^{-l-1}  \right] P_{l}^m(\cos \theta)\cos(m \phi)
\\
\label{eq:general-solutions_n}
n &=&\sum_{l=0}^{\infty}\sum_{m=0}^{l} \left[
C_{lm} j_l(\kappa r)+ D_{lm} n_l(\kappa r)
\right] P_{l}^m(\cos \theta)\cos(m \phi).
\end{eqnarray}
Here, $P_{l}^m(x)$ are the associated Legendre polynomials, and $j_l,n_l$ are the $l$-th order spherical Bessel functions of the first and second kind~\cite{Jackson1998,Morse1953}. In addition, $\kappa=\sqrt{i\omega/D}=(1+i)\sqrt{\omega/(2D)}$ is a complex 
wave vector. Here, we have chosen a coordinate system such that the solutions of $\psi$ and $n$ are an even function of $\phi$, i.e., the reference axis for the azimuthal angle $\psi$ is along one of the principle axes of the simple cubic lattice~\cite{McPhedran2}.

Unlike in the case of a single sphere, the expansion of the general solution in Eq.~\eqref{eq:general-solutions_psi} contains terms of the form $r^l$ which diverge at infinity. Because we are using a lattice with multiple centers, these terms are incoming waves which arise from the scattered waves leaving the other centers, in the framework of multiple scattering theory. This physical picture will prove to be the key to solve for $\psi$ in a periodic lattice~\cite{Rayleigh1892}. Similarly, both the $j_l$ and $n_l$ terms diverge at infinity in Eq.~\eqref{eq:general-solutions_n}.  Again, an understanding of the multiple scattering nature of the system is required for us to proceed~\cite{Korringa,KohnRostoker}. 

Considerable simplification can be achieved by carefully taking into account the symmetry. For a simple cubic lattice with $\vec{E}$ aligned along its major axis, 
the potential must change signs when $\theta$ goes to $\pi-\theta$. As a result, only odd integers $l$ are allowed in the solutions. Because a rotation of $\pi/2$ around the azimuthal axis leaves the system invariant, only $m$'s that are divisible by four are allowed in both expansions. Furthermore, becauase the crystal lattice breaks the rotational symmetry, unlike in the case of a spherically or axially symmetric single scatterer (e.g. a sphere or a cylinder), $m\neq0$ terms in both expansions cannot be ignored. We will see explicitly in the next section that the structural constants derived from the crystal structures couple different angular momentum states.

By inserting Eqs.~\eqref{eq:general-solutions_psi} and~\eqref{eq:general-solutions_n} into the boundary conditions in Eq.~\eqref{eq:BC-explicit} at the outer surface of the double layer, 
$r=a+\delta\approx a$, using $L^2 P_{l}^{m} (\cos \theta) \cos (m \phi)=-l(l+1) P_{l}^{m}(\cos \theta) \cos (m \phi)$, and matching the coefficients of the orthogonal basis sets $P_{l}^{m}(\cos \theta) \cos (m \phi)$, we have the following coupled linear equations relating the 
coefficients $A_{lm}$, $B_{lm}$, $C_{lm}$, and $D_{lm}$:
\begin{widetext}
\begin{equation}
\label{eq:BC-matched}
\begin{split}
\kappa\left[C_{lm}j_l'(\kappa a)+ D_{lm}n'_l(\kappa a)\right]&- \left[l A_{lm} a^{l-1}-(l+1) B_{lm} a^{-l-2}\right]=0,\\
- 2\kappa\left[
C_{lm}j_l'(\kappa a)+ D_{lm}n'_l(\kappa a)\right]&+
\frac{\Omega}{N_0}\frac{l(l+1)}{a^2}\left[\left( A_{lm} a^l + B_{lm} a^{-l-1} \right) + C_{lm} j_l(\kappa a)+ D_{lm} n_l(\kappa a)
\right]=0,
\end{split}
\end{equation}
\end{widetext}
Here we have absorbed the factor $\left(e/k_B T\right)$ into the definition of $\psi$ and have rescaled $n(\vec{r})\to N_0 n(\vec{r})$. The primed functions $n'_l$ and $j'_l$ are the derivatives of the spherical Bessel functions. The matched boundary conditions in Eq.~\eqref{eq:BC-matched} do not mix terms with different $\langle l,m\rangle$ indexes.

It is worthwhile to discuss some physical implications of the governing equations in Eq.~\eqref{eq:governing-beyond-DL} and the boundary conditions in Eq.~\eqref{eq:BC-matched}. First, when combining Eq.~\eqref{eq:governing-beyond-DL} and Eq.~\eqref{eq:BC-matched} with the two extra boundary conditions, $\psi\to -H z$ and $n \to 0$ at $|\vec{r}|\to \infty$, which are applicable for a single sphere, one can show that only the $\ell=1$ term is non-vanishing and derive the single-particle polarization response for a charged sphere immersed in the electrolyte, given by~\cite{Chew1982,Chassagne2008,Hou2017}:
\begin{equation}
\label{eq:polarization_coefficient-single_particle}
P = \frac{\xi - \left( 1 - \frac{\xi}{\kappa a} \frac{ h_1^{(1)}(\kappa a) }{  h'^{(1)}_1 (\kappa a) }  \right)  }{ \xi  +  2  \left( 1 - \frac{\xi}{\kappa a} \frac{ h_1^{(1)}(\kappa a) }{ h'^{(1)}_1 (\kappa a) }  \right) },
\end{equation}
where $\xi\equiv \Omega/(N_0 a)$ can be viewed as the effective ratio of the particle conductivity to the electrolyte conductivity. Also, $h^{(1),(2)}_{\ell}\equiv j_{\ell} \pm i n_{\ell}$ are the spherical Hankel's function of the first and second kinds~\cite{Morse1953}, respectively, and $h'^{(1),(2)}_{\ell}$ are their derivatives. This single particle polarization response is often the starting point for constructing effective medium models to describing the response of the complex conductivity~\cite{Sihvola1999}. As the baseline for our following discussions, we will use this result together with the Maxwell Garnett mixing formula to obtain a prediction for the dielectric response. Typically, this approach is valid for low volume fractions of the spheres and, as we will show later, at higher frequencies. 

In addition,  the Helmholtz equation for the perturbed ion density in Eq.~\eqref{eq:governing-beyond-DL} indicates that there is a length scale for the decay given by $\sqrt{D/\omega}$. Hence, the length scale becomes longer at lower frequencies. Given that $D \sim 10^{-9}$ m$^2$/s, the decay length scale is on the order of microns for $\omega\sim 10^3$ radians/s. This implies that the presence of obstacles or other charged spheres within the length scale of the decay will interfere with the ion redistribution due to the applied electric-field. Hence, we expect that the polarization response of densely packed charged spheres will exhibit different behavior compared to that for a single charged sphere in Eq.~\eqref{eq:polarization_coefficient-single_particle} iat low frequency. Periodic boundary conditions, which will be derived in the next section, will allow us to account for the interference of the polarization response between different charged spheres.

\section{Periodic boundary conditions}
\label{sec:Periodic_BCs}

We will now derive the corresponding periodic boundary conditions for both Laplace's equation, satisfied by the electric potential, and the Helmholtz equation, satisfied by the ion densities, for a periodic arrangement of charged spheres. It is useful to recognize that the two conditions are independent of each other. All the interactions between the potential and the ion density are encoded in the boundary conditions discussed in the previous section. However, because the presence of the lattice structure breaks the spherical symmetry, we expect that the periodic conditions will mix terms with different $\langle l,m\rangle$ indexes.

\subsection{Periodic conditions for the Laplace Equation}
\label{Laplace}

The conceptually simpler of the two conditions is the one for the Laplace equations in a periodic array for $\psi$. As illustrated by Lord Rayleigh~\cite{Rayleigh1892}, insight can be gained by carefully studying the structure of the general solution given in Eq.~\eqref{eq:general-solutions_psi}. The potential $\psi$ in the space outside the sphere at origin $O$ has three contributions: (i) the part that comes from the charges on the surface of this sphere (its double-layer etc.); (ii) the part contributed by the charges of all other spheres and (iii) the external driving potential $-\vec{E}\cdot{r}$. Part (i), and only this part, will decay monotonically as $r$ increases.  This is represented by the \emph{decreasing} parts, the $r^{-l-1}$ terms, in the general solution evaluated point $Q=(r,\theta,\phi)$, given in Eq.~\eqref{eq:general-solutions_psi}. Conceptually, these are equivalent to the ``outgoing waves'' of $\psi$ with the sphere at the origin $O$ as the ``source''.

The ``incoming waves'', consisting of the $r^l$ terms in the general solution, must be precisely contributed from (ii) the surface charge of all the \emph{other} spheres and (iii) the driving potential. However, from the point of view of the \emph{other} spheres $i$, what it is contributing is precisely \emph{its} outgoing wave, of the form 
$\displaystyle \sum_{lm} B_{lm}\rho_i^{-l-1} P_{l}^m(\cos \theta_i) \cos(m \phi_i)$, where $(\rho_i,\theta_i,\phi_i)$ are the coordinates of the point $Q$ above, but in the spherical coordinate system centered at the $i$-th sphere. Here, we have used the fact that the lattice is periodic, which means that the spheres' $A_{lm}$ and $B_{lm}$ coefficients are the same.

With these observations, we have a powerful identity, valid for any point $Q$.  (A more complete proof than the argument given here can be found in Sec. 2 of Ref.~\cite{McPhedran2}):
\begin{equation}
\label{eq:phi-matching}
\begin{split}
&\sum_{l=1}^{\infty}\sum_{m=0}^{2l-1} A_{2l-1,m} r^{2l-1} P_{2l-1}^{m}(\cos\theta) cos(m\phi)
\\
=& \sum_{i\neq 0}\sum_{l=1}^{\infty}\sum_{m=0}^{2l-1}  B_{lm} \rho_i^{-2l} P_{2l-1}^{m}(\cos \theta_i) cos(m\phi_i)
	-  H x.
\end{split}
\end{equation}
Here, we use the form of the expansion in Eq.~\eqref{eq:general-solutions_psi} and take into account that only terms with odd integer $l$ have non-vanishing coefficients due to the symmetry. Again, $(r,\theta,\phi)$ is the coordinate of point $Q$ in the polar coordinate system centered at the origin $O$, whereas $(\rho_i,\theta_i,\phi_i)$ is the same point represented in coordinates centered around the sphere $i$ \emph{not} at the origin: $\vec{\rho}_i=\vec{r}-\vec{R}_i$, where
$\vec{R_i}=(R_i,\Theta_i,\Phi_i)$ is a nonzero Bravais lattice vector.

The most straightforward way to extract useful equations relating $A_{lm}$ and $B_{lm}$ from Eq.~\eqref{eq:phi-matching} is to use a generalization of the addition theorem, c.f. Ref.~\cite{McPhedran2}, to transform between representations in polar coordinate systems centered around different origins:
\begin{equation}
\label{eq:addition}
\begin{split}
\frac{P_{l}^{m}(\cos\theta_i)e^{i m \phi_i } }{\rho_i^{l+1}}=&
\sum_{l'm'}(-1)^{l+m'}\frac{(l+l'+m'-m)!}{(l-m)!(l'+m')!}
\frac{r^{l'}}{R_i^{l+l'+1}}
\\
& \times P_{l'}^{m'}(\cos\theta)P_{l+l'}^{m-m'}(\cos \Theta_i) e^{i m' \phi} e^{ i(m-m')\Phi_i}.
\end{split}
\end{equation}
Putting Eq.~\eqref{eq:addition} back into Eq.~\eqref{eq:phi-matching} and using the orthogonality of the Legendre polynomials, we obtain the following equations for $A_{lm},B_{lm}$ (c.f. Eq. 16 of ~\cite{McPhedran2}):
\begin{widetext}
\begin{equation}
\label{eq:psi-final}
	-H \delta_{l1}\delta_{m0}=\sum_{l'=1}^{\infty} \sideset{}{'}\sum_{m'=0}^{2l'-1}\bigg\{
	\delta_{ll'}\delta_{mm'}\frac{(2l-1+m)!}{(2l-1-m)!\epsilon_m}A_{2l'-1,m'}
	+\frac{(-1)^{m_{<}} U_{2(l+l'-1)}^{|m'-m|}+U_{2(l+l'-1)}^{m+m'} }{2(2l-1-m)!(2l'-1-m')!} B_{2l'-1,m'}
	\bigg\},
\end{equation}
\end{widetext}
where $m_<$ is the smaller of $m,m'$; $\epsilon_m=1$ for $m=0$ and 2 otherwise; and the prime in the summation indicates that only those $m$ divisible by $4$ are included. The $U_l^m$ are a series of \emph{geometric} factors, similar to the ``structure factors'' $G_{l,m;l',m'}$ below in the KKR calculations and are defined as:
\begin{equation}
\label{eq:phi-structure}
	U_l^m=(l-m)!\sum_{i\neq 0} 
	R_i^{-l-1}P_l^{m}(\cos(\Theta_i))\cos(m\Phi_i).
\end{equation}
Once these sums, over all non-zero Bravais vectors $R_i$, are computed and tabulated~\cite{McPhedran1}, all the coefficients in the linear equations~\eqref{eq:psi-final} are determined.

\subsection{Periodic conditions for the Helmholtz Equation}
\label{Helmholtz}

In this section, we will use the periodicity of the lattice to derive an additional relation between the coefficients $C_{lm}$ and $D_{lm}$ that appear in the expansion for the perturbed ion density. To obtain this relation, we will make use of a formulation adapted from the KKR method (described more fully below.) This method requires only that we can express our problem in terms of a periodic potential that is non-zero only in a localized region inside each unit cell. In our case, this region can be taken to be the space occupied by the spherical particle and its double layer. As we will show below, only the existence of this effective (or ``fictitious'') potential is required. The actual form of this potential does not matter and does not need to be considered.

To solve the Helmholtz equation for the perturbed ion density $n$ in the charge-neutral liquid outside the double layer, we need to take full advantage of the periodicity of the lattice. Because the boundary conditions in Eq.~\eqref{eq:BC-explicit} involve only derivatives of the potential $\psi$, and the derivative of the potential from the driving force $\vec{E}\cdot\vec{r}$ has no spatial dependence, the ion concentration $n$ is indeed translationally invariant with respect to the Bravais lattice vectors. Hence, our solution should satisfy Bloch's theorem, given by
\begin{equation}\label{eq:bloch}
n_{\vec{k}}(\vec{r}+\vec{R}_i)= 
n_{\vec{k}}(\vec{r})\exp(i\vec{k}\cdot\vec{R}_i)
\end{equation}
for all Bravais lattice vectors $\vec{R}_i$. This reduces all the physics to a single primitive cell.

To proceed, we now adapt the KKR method~\cite{Korringa,KohnRostoker}, which deals with the so-called ``muffin-tin'' potential: a localized potential that is non-vanishing only in a small sphere at the center of the unit cell. Although our problem does not really have such a potential, we can justify the existence of an effective potential of that form as follows: The boundary conditions in Eq.~\eqref{eq:BC-matched} will cause a complex phase shift for each partial wave component at the boundary of the double layer. As can be seen in Eq.\eqref{eq:general-solutions_n}, the contribution for each $l$ and $m$ can be rewritten as a sum of an incoming and outgoing 1D wave, and the ratio between the two amplitudes gives the complex phase shift. For each of these partial wave components, the Helmholtz equation thus reduces to a 1D radial problem (c.f. Ref.~\cite{Landau1965} \S 32), and one can always find an equivalent potential that creates the same phase shifts (\emph{ibid.},\S 21). Because the waves are freely propagating, except at the boundary of the double layer, the equivalent potential can always be chosen so that it vanishes outside of the double layer. Note that when the partial waves are recombined, the full equivalent potential $V(\vec{r})$ is generally no longer spherically symmetric, but is still confined within the sphere. For simplicity of notation, we will use an effective ``Schr\"odinger equation'' with an equivalent potential $V$ in our formal derivations below. More accurately, the potential $V$ should be thought of as an abstract operator encoding boundary changes that is spatially confined within the sphere and the thin double layer.

Here, we will derive our formalism in a slightly different way from the original variational approach used by the KKR method~\cite{KohnRostoker,Korringa}. As hinted in the appendix in Kohn and Rostoker's original paper, we will use the Green's function and integral equation formalism, although the basic physics is that of multiple scattering. It is worth noting that our derivations do not depend on the specific form of $V$ other than that it vanishes outside the sphere and double layer, and we are able to arrive at the final coupled equations for the coefficients $C_{lm}, and D_{lm}$.

Let us start with the Green's function for the wave equation in free space:
\begin{equation}
\label{eq:green-def}
	(\nabla^2+E)G_{E}(\vec{r},\vec{r}\,')
	=\delta(\vec{r}-\vec{r}\,'),
	\quad
	G_{E}(\vec{r},\vec{r}\,')= -
	\frac{e^{i\kappa|\vec{r}-\vec{r}\,'|}}{4\pi(|\vec{r}-\vec{r}\,'|)}.
\end{equation}
Here $E \equiv i\omega/D =\kappa^2$ to be consistent with Helmholtz equation of interest defined in Eq.~\eqref{eq:governing-beyond-DL}. As discussed above, we convert the boundary conditions into a fictitious short-range single-site potential $V(\vec{r})$ confined within the sphere. Effectively, our problem is described by the following equations:
\begin{equation}
\label{eq:integral-formulation}
\begin{split}
	&-\nabla^2 n(\vec{r})+ U(\vec{r}) n(\vec{r})=E 
 n(\vec{r}),
 \\
	& n(\vec{r})=\int~G_{E}(\vec{r},\vec{r}\,') U(\vec{r}\,') 
	n(\vec{r}\,') d^3\vec{r}\,',
\end{split}
\end{equation}
where the integral is over the entire space and $U$ is the single-site function $V$ translated to every site in the lattice of spheres. Thus $U$ is the true ``potential'' of the lattice. The second equation is simply the integral form of the first equation given Eq.~\eqref{eq:green-def}. This can be verified by applying the operator $\nabla^2+E$ to both sides of the integral equation, where the $\nabla$ acts on the vector $\vec{r}$.  

We now break up the integral over all space in Eq.~\eqref{eq:integral-formulation} into a summation over each unit cell and then make a change of variable $\vec{r}\,''=\vec{r}\,'-R_i$ in each unit cell.  This translates the integral back to the unit cell containing the origin of our polar coordinates. Because the $R_i$ are Bravais lattice vectors, we have, by Bloch's theorem, $n_{\vec{k}}(\vec{r}+\vec{R})= n_{\vec{k}}(\vec{r})\exp(i\vec{k}\cdot\vec{R})$, whereas the translated part of $U$ becomes $V$ in the unit cell at the origin. Finally, the integral equation becomes (with $r''\to r'$ at the end):
\begin{equation}
\label{eq:bloch-integral-eq}
\begin{split}
	n(\vec{r})&=\int_D~\mathcal{G}_{E,\vec{k}}(\vec{r},\vec{r}\,') 
	V(\vec{r}\,') n(\vec{r}\,') d^3\vec{r}\,',
	\\
	\mathcal{G}_{E,\vec{k}}(\vec{r},\vec{r}\,')&=\sum_{i}
	G_{E}(\vec{r},\vec{r}\,'+\vec{R}_i) e^{i\vec{k}\cdot\vec{R}_i}.
\end{split}
\end{equation}
Here, the domain of the integral, $D$, covers only the single unit cell at the origin. The summation is over all Bravais lattice points and encodes \emph{all} the information of Bloch's theorem into a modified Green's function $\mathcal{G}_{E,\vec{k}}$, which depends on an undetermined wave-vector, $\vec{k}$.

In the usual band-structure computation, the relation between $E=\kappa^2$ and $\vec{k}$ is simply the dispersion relation and is obtained from a secular equation. However, the presence of the non-homogeneous equation~\eqref{eq:psi-final} (the $-H$ term for $l=1,m=0$) coupled to the rest of boundary conditions~\eqref{eq:BC-matched} separates our case from usual quantum mechanics calculations. This makes physical sense because the size of $n$ should be set by the strength of driving field $H$. Also, unlike in quantum mechanics, where the periodicity of the expectation value over the complex wavefunction is not affected by a Bloch phase factor, the ion density $n(\omega)$ is a ``true'' physical observable with its periodicity directly affected by the Bloch vector $\vec{k}$. We thus need to honor this periodicity, except when there is mechanism to break it.  Because we are interested in the linear response regime with respect to the driving electric field, we shall require our solution to preserve the spatial periodicity. In this spirit, we can only have $\vec{k}=0$, which is consistent with the condition used in Refs.~\onlinecite{SenKan1,SenKan2} at the frequency $\omega=0$. Hereafter, we will drop all the $\vec{k}$ dependence with the understanding that we have already taken $\vec{k}=0$.

The confinement of the ``potential'' or operator $V$ within a sphere (the original sphere plus a thin double layer) brings a further simplification: the domain, $D$, of the integral in 
Eq.~\eqref{eq:bloch-integral-eq} is in fact the sphere with $r\le a+\delta\approx a$. Let us consider two points, $\vec{r}$ and $\vec{r}\,'$, within domain $D$. First, we observe that, analogous to Eq.~\eqref{eq:green-def}, $(\nabla'^2+E)\mathcal{G}_{E}(\vec{r},\vec{r}\,')=\delta (\vec{r}-\vec{r}\,')$, where the prime on the $\nabla$ indicates that the derivative is with respect to the vector $\vec{r}'$. This is because only the first term, with $R=0$ in the summation in Eq.~\eqref{eq:bloch-integral-eq}, is relevant for $r$ and $r'\le a$. Multiplying both sides by $n(\vec{r}\,')$ and integrating over the unit cell, $D$, we have
\begin{equation}\label{eq:Green-derive1}
	n(\vec{r}\,)=\int_D d\vec{r}\,'
	n(\vec{r}\,')(\nabla'\,^2+E)\mathcal{G}_{E}(\vec{r},\vec{r}\,').
\end{equation}
On the other hand, we have $(\nabla'^2+E) n(\vec{r}\,')=V(\vec{r}\,') n(\vec{r}\,')$ within domain $D$. Multiplying both sides by $\mathcal{G}_{E}$, integrating both sides over $D$, and then using Eq.~\eqref{eq:bloch-integral-eq}, we have:
\begin{equation}\label{eq:Green-derive2}
	n(\vec{r}\,)=\int_D d\vec{r}\,'
	\mathcal{G}_{E}(\vec{r},\vec{r}\,')(\nabla'\,^2+E)n(\vec{r}\,').
\end{equation}
By subtracting Eq.~\eqref{eq:Green-derive1} from Eq.~\eqref{eq:Green-derive2}, and using the identity 
$A\overrightarrow{\nabla}\cdot\overrightarrow{\nabla}B=
\overrightarrow{\nabla}\cdot(A\overrightarrow{\nabla}B)-
\overrightarrow{\nabla}A\cdot\overrightarrow{\nabla}B$, we can convert the volume integral into a \emph{surface} integral, evaluated at the outer edge of the double layer $\Sigma$:
\begin{equation}\label{eq:surface-integral}
	\int_{\Sigma} dS' \left[
	n(\vec{r}\,')\frac{\partial}{\partial r'} 
 \mathcal{G}(\vec{r},\vec{r}\,')-
 \mathcal{G}(\vec{r},\vec{r}\,')
\frac{\partial}{\partial r'}n(\vec{r}\,')
 \right]=0
\end{equation}
This integral equation on the outer \emph{surface} of the double layer encodes \emph{all} the physics, from the periodic structure to the boundary conditions, throughout the \emph{entire space}~\footnote{The derivation here is not entirely rigorous. The Green's function $\mathcal{G}_{E,\vec{k}}(\vec{r},\vec{r}\,')$ becomes singular as $\vec{r}$ and $\vec{r}\,'$ coincide, so the integrals in Eqs.~\eqref{eq:Green-derive1} and~\eqref{eq:Green-derive2} require more care, i.e., by restricting the range of integration to near, but not at, the surfaces of the spheres, c.f.~\cite{KohnRostoker} Appendix I for more details}.

The general solution for $n$ in Eq.~\eqref{eq:general-solutions_n} provides an explicit expression at the surface $\Sigma$ in terms of the associated Legendre polynomials.  An expansion of $\mathcal{G}$ can be derived from the addition theorem for the Legendre polynomials, c.f. Ref.~\cite{KohnRostoker}, Appendix II, given by
\begin{widetext}
\begin{equation}
\label{eq:Green-expansion}
	\mathcal{G}_{E}(\vec{r},\vec{r}\,')=\sum_{l,m}\sum_{l',m'}
	\left[G_{lm;l'm'} (\kappa)j_l(\kappa r)j_{l'}(\kappa r')+
	\kappa\delta_{l,l'}\delta_{m,m'}j_l(\kappa r)n_{l}(\kappa r')
	\right]Y_{lm}(\theta,\phi)Y^*_{l'm'}(\theta',\phi'),
\end{equation}
\end{widetext}
where $Y_{lm}(\theta,\phi)$ are the spherical harmonics which can be easily related to the associated Legendre polynomials~\cite{Jackson1998}. Also, $G_{lm,l'm'}(\kappa)$ are so-called ``structure factors'' (denoted as $A_{lm,l'm'}$ in Ref.~\cite{KohnRostoker}) and depend only on $E$ and the lattice structure. The functional forms for the structure factors are given by~\cite{KohnRostoker}
\begin{widetext}
\begin{equation}
\label{eq:structure-factors}
G_{l,m;l',m'}(\kappa)= -  \frac{ (4\pi)^2 i^{(l-l')}}{ \tau j_{l} (\kappa r) j_{l'} (\kappa r')}
\sum_{n} \frac{j_l(|\vec{K}_n|r)j_{l'}(|\vec{K}_n|r') Y_{lm}^*(\theta_n,\phi_n) Y_{l'm'}(\theta_n,\phi_n)}{(\vec{K}_n+\vec{k})^2-E}
-\kappa \delta_{ll'} \delta_{mm'} \frac{n_l(\kappa r')}{j_l(\kappa r')}
\end{equation}
\end{widetext}
Here, $\tau$ is the volume of the unit cell. The $\vec{K}_n$ are the vectors of the reciprocal lattice, and $\theta_n$ and $\phi_n$ are the polar angles of $\vec{K}_n$ relative to the origin. For the simple cubic lattice, we have $\tau=b^3$ and $\vec{K}_n=(2\pi/b) \left( n_x \hat{x} + n_y \hat{y} + n_z \hat{z}\right)$ for integer $n_{x,y,z}$. Values of $G_{l,m;l',m'}(\kappa)$ for real $E$ have been extensively tabulated for KKR calculations~\cite{Davis1971}. However, in our problem, $E=\kappa^2=i\omega/D$, an imaginary number, which means it is outside the typical region of interest for the KKR calculations. Hence, we need to evaluate the slowly converging sum in Eq.~\eqref{eq:structure-factors}. In Appendix~\ref{app:structure-factor}, we will review an efficient method for computing the structure factor developed by Ham and Segall~\cite{Ham1961}.

By substituting Eq.~\eqref{eq:Green-expansion} and the general solution for $n$ in Eq.~\eqref{eq:general-solutions_n} into the surface integral equation, Eq.~\eqref{eq:surface-integral}, and carrying out the angular integration using the orthogonality of the functions $P_{l}^{m}(\cos \theta') \cos(\phi')$, we  obtain linear equations involving 
$C_{lm},D_{lm}$ given by
\begin{widetext}
\begin{equation}
\label{eq:n-final}
\begin{split}
&- \kappa \frac{1}{\epsilon_m \sqrt{4l-1}}  \frac{\sqrt{(2l-1+m)!}}{\sqrt{(2l-1-m)!}} C_{2l-1,m}
\\
+& \sum_{l'=1}^{\infty} \sum_{m'=0}^{2l'-1} \frac{1}{2 \sqrt{4 l' -1}} \frac{\sqrt{(2l'-1+m')!}}{\sqrt{(2l'-1-m')!}} \left[G_{2l-1,m;2l'-1,m'}+(-1)^{m'} G_{2l-1,m;2l'-1,-m'}  \right] D_{2l'-1,m'}
=0.
\end{split}
\end{equation} 
\end{widetext}
Again, we have used the fact that all the $C_{lm}$ and $D_{lm}$ with even index $l$ vanish, and that $\epsilon_m=1$ for $m=0$ and is 2 otherwise.

Finally, we have a set of \emph{linear} equations for the coefficients $A_{l,m},B_{l,m},C_{l,m}$, and $D_{l,m}$ given by Eqs.~\eqref{eq:BC-matched},~\eqref{eq:psi-final}, and~\eqref{eq:n-final}. Given an angular momentum cutoff $\{l\le L, m\le M\}$, there are exactly the same number of unknowns and equations. As a result, we can, in principle, evaluate these coefficients to arbitrary precision by increasing the values of the cutoff $\langle L, M \rangle$. In the next section, we will show that only $B_{1,0}$ is needed to obtain the complex conductivity (or dielectric constant) of the system. 

\section{Complex Conductivity: relation to $B_{1,0}$ and the Lowest order formula}
\label{sec:conductivity_B10}

In this section, we first follow Lord Rayleigh~\cite{Rayleigh1892} by relating the value of the complex conductivity to the coefficient $B_{1,0}$ in Eq.~\eqref{eq:general-solutions_psi}. With this relation, we derive the lowest order formula for the complex conductivity. We will then compare this derived formula to that of the single-particle polarization response from the Maxwell Garnett mixing formula. 

The basis of Lord Rayleigh's trick starts with the following form of Green's theorem:
\begin{equation}
\label{eq:green}
\varoiint_{\Sigma} dS \left(
\phi~\hat{n}\cdot\overrightarrow{\nabla}\psi-
\psi~\hat{n}\cdot\overrightarrow{\nabla}\phi
\right)=0,
\end{equation}
where $\Sigma$ is the surface of a singly connected region $D$, $\hat n$ is the unit vector normal to this surface, and $\psi$ and $\phi$ are potentials obeying  Laplace's equation. We now choose $D$ to be the region between the sphere and the boundary of the unit cell and take $\psi$ to be our potential in Eq.~\eqref{eq:general-solutions_psi}. Assuming the external field $\vec{E}$ is along the $z$ direction, we take $\phi =z = r\cos\theta$.

To evaluate the integral, we begin with the four surfaces of the unit cell that are parallel to the z direction. The electric field $\overrightarrow{\nabla}\psi$ and the gradient of $\phi$ at each of these four surfaces are  parallel to the surfaces due to the symmetry of the simple cubic lattice. Thus, the surface integral vanishes on these four sides. Next, $\overrightarrow{\nabla}\psi$ evaluated at the other two sides of the unit cell  is perpendicular to these two sides.  Its magnitude is $-E_z(x,y)|_{z=\pm b/2}=-j_{z}(x,y)/\sigma_w$ because the medium there is simply the interstitial electrolyte solution outside the double layer, with conductivity $\sigma_w(\omega)$. Here, we have use the periodic condition $E_z(x,y)|_{z=+ b/2}= E_z(x,y)|_{z=- b/2}$. Then the first term of the surface integral in Eq.~\eqref{eq:green} simply give $b^3\langle j_{z} \rangle / \sigma_w$ with $\langle j_{z} \rangle $ defined as surface averaged current density at the cell boundary normal to the applied electric field.

The second term of the integrand evaluated at these two surfaces is simply the drop in potential $\psi$ between the two sides. By the translational symmetry in the 
z-direction of the lattice and  perturbed charges, this potential drop is simply $H b$~\cite{McPhedran1}. Finally, the contribution from the inner sphere surface can be computed because the $\cos(\theta)=P_1(\cos\theta)$ angular dependence of $\phi$ and $\psi$ leaves only the $\langle l=1,m=0 \rangle$ term non-vanishing. Putting all these together, we can write the complex conductivity, $\sigma_s$, of the mixture as
\begin{equation}\label{eq:conductivity}
\sigma_s=\frac{\langle j_z\rangle}{H}=\sigma_w(\omega)\left(1+\frac{4\pi B_{1,0}}{H b^3}\right).
\end{equation}
Here, $\sigma_w(\omega) =\sigma_w'(\omega) - i \varepsilon_0 \omega \epsilon_w'(\omega)$ is the frequency-dependent conductivity of the electrolyte solution, where $\epsilon_w'(\omega)$ and $\sigma_w'(\omega)$ are the relative permittivity and conductivity of the electrolyte. As usual, $\varepsilon_0$ is the vacuum permittivity. Thus, to get the complex conductivity of charged spheres arranged in simple cubic lattice immersed in the electrolyte solution, one only needs the single coefficient $B_{1,0}$ in the solution Eq.~\eqref{eq:general-solutions_psi}. 

To further simplify our computations, it is convenient to rescale all quantities with respect to the lattice spacing $b$. Namely, we shall use the resacled variables (denoted by the overhead bar symbol) given by $\bar{r} = r/b$, $\bar{a}=a/b$, $\bar{N}_0=b^3 N_0$, $\bar{\Omega}=b^2 \Omega$, $\bar{\kappa}=b \kappa$, $\bar{E}= b^2 E$, and $\bar{H}=Hb$. The reciprocal lattice vectors are also defined accordingly as $\vec{\bar{K}}_n=2\pi \left( n_x \hat{x} + n_y \hat{y} + n_z \hat{z}\right)$ in the rescaled coordinates. The rescaled expansion coefficients are defined as $\bar{A}_{l,m}=b^l A_{l,m}$, $\bar{B}_{l,m}=B_{l,m}/b^{l+1}$, $\bar{C}_{l,m}=C_{l,m}$ and $\bar{D}_{l,m}=D_{l,m}$. At the same time, the two structure factors becomes $\bar{U}_{l}^{m} = b^{l+1} U_{l}^{m}$ and $\bar{G}_{l,m;l',m'}(\kappa) = b G_{l,m;l',m'}(\kappa)$. In terms of the rescaled variables, we have
\begin{equation}
\label{eq:conductivity_rescaled}
\sigma_s=\sigma_w(\omega)\left(1+\frac{4\pi \bar{B}_{1,0}}{\bar{H}}\right).
\end{equation}
Hereafter, we will always use the rescaled quantities.

To gain further intuition, it is useful to obtain the analytic form of $\bar{B}_{1,0}$ and, hence, the complex conductivity, to lowest order. With the angular 
momentum cutoff $\langle L=1, M=0 \rangle$, Eqs.~\eqref{eq:BC-matched},~\eqref{eq:psi-final} and~\eqref{eq:n-final} give four coupled linear equations for four variables. After some algebra, we obtain the following prediction for the response of the conductivity to the $\langle L=1, M=0 \rangle$ order:
\begin{equation}
\label{eq:conductivity-10_order}
\frac{\sigma^{\langle 1, 0 \rangle}_s}{\sigma_w(\omega) }= 1 + f  \frac{\xi - \left( 1 - \frac{\xi}{\bar{\kappa} \bar{a}} \frac{ \bar{G}_{1,0;1,0} j_1(\bar{\kappa} \bar{a})+ \bar{\kappa} n_1(\bar{\kappa} \bar{a})  }{  \bar{G}_{1,0;1,0} j'_1(\bar{\kappa} \bar{a})+ \bar{\kappa} n'_1(\bar{\kappa} \bar{a}) }  \right)  }{  \frac{1- f}{3} \xi +  \frac{2+ f}{3}  \left( 1 -  \frac{\xi}{\bar{\kappa} \bar{a}} \frac{ \bar{G}_{1,0;1,0} j_1(\bar{\kappa} \bar{a})+ \bar{\kappa} n_1(\bar{\kappa} \bar{a})  }{  \bar{G}_{1,0;1,0} j'_1(\bar{\kappa} \bar{a})+ \bar{\kappa} n'_1(\bar{\kappa} \bar{a}) }  \right) } ,
\end{equation}
where, $\xi = \Omega/(N_0 a)= \bar{\Omega}/(\bar{N}_0 \bar{a}) $ and $f= 4 \pi \bar{a}^3/3$ is the volume fraction of charged spheres. We have used $U_{2}^{0}=4\pi/3$ as argued by Lord Rayleigh for a needle-shaped sample~\cite{Rayleigh1892,McPhedran1}. The method for evaluating $\bar{G}_{1,0;1,0}(\bar{\kappa})$ is discussed in Appendix~\ref{app:structure-factor}. Interestingly, one can numerically confirm that $\bar{G}_{1,0;1,0}(\bar{\kappa}) \to -i \bar \kappa$ when $|\bar{\kappa}| \to \infty$, and $\bar{G}_{1,0;1,0}(\bar{\kappa}) \bar{\kappa}^2 \to 4\pi$ when $|\bar{\kappa}| \to 0$. The latter limit allows us to show that solution in Eq.~\eqref{eq:conductivity-10_order} is consistent with that found in Ref.~\onlinecite{Sen1987} to comparable order.

It is useful to compare this lowest order result with that from the Maxwell Garnett mixing formula using the polarization response of a single particle.  The Maxwell Garnett mixing formula can be expressed as~\cite{Garnett1904,Garnett1906}:
\begin{equation}
\sigma_{\rm MG} =\sigma_{w}(\omega) \left(1 + 3 f \frac{P}{1-f P}\right),
\end{equation}
where $f= 4 \pi \bar{a}^3/3$ is simply the volume fraction of charged spheres. By using the single-particle polarization coefficients in Eq.~\eqref{eq:polarization_coefficient-single_particle}, we find that the complex conductivity predicted by the Maxwell Garnett formula is given by
\begin{equation}
\label{eq:M-G-conductivity}
\frac{\sigma_{\rm MG\_SP}}{\sigma_{w} } =1 + f \frac{\xi - \left( 1 - \frac{\xi}{\bar{\kappa} \bar{a}} \frac{ h_1^{(1)}(\bar{\kappa} \bar{a}) }{  h'^{(1)}_1 (\bar{\kappa} \bar{a}) } \right)  }{ \frac{1- f}{3} \xi  +  \frac{2+ f}{3}  \left( 1 - \frac{\xi}{\bar{\kappa} \bar{a}} \frac{ h_1^{(1)}(\bar{\kappa} \bar{a}) }{ h'^{(1)}_1 (\bar{\kappa} \bar{a}) }  \right) } .
\end{equation}
We observe that Eqs.~\eqref{eq:conductivity-10_order} and~\eqref{eq:M-G-conductivity} have a similar functional form. 

Remarkably, in the limit as $|\bar{\kappa}| \to \infty$, one can verify that $\sigma^{\langle 1, 0 \rangle}_s \to \sigma_{\rm MG}$ by using relations between spherical Hankel's and Bessel's functions~\cite{Morse1953}. Recalling that $\bar{\kappa} = b \sqrt{i\omega/D}$, this implies that $\sigma_{\rm MG}$ becomes valid: (1) for dilute mixtures with $b\gg \sqrt{D/\omega}$, or equivalently, (2) in the high frequency regime, with $\omega \gg D/ b^2$. As expected, the Maxwell Garnett mixing formula should capture the response in the dilute limit. Our result here also reveals a less intuitive frequency dependence, that the Maxwell Garnett mixing formula becomes more valid at higher frequencies.

\section{Complex Conductivity: Higher order evaluations}
\label{sec:conductivity_higher_order}

In this section, we will first recast the four sets of independent linear equations in Eq.~\eqref{eq:BC-matched},~\eqref{eq:psi-final} and~\eqref{eq:n-final} into two sets of equations. This allows us to effectively set up numerical calculations for higher angular momentum cutoffs. Let us start from Eq.~\eqref{eq:psi-final} and write $\bar{A}_{2l+1,m}$ explicitly as a summation over the $\bar{B}_{2l+1,m}$, as follows:
\begin{equation}
\label{eq:A_in_B}
\begin{split}
& \bar{A}_{2l-1,m} =-\bar{H} \delta_{l1}\delta_{m0} 
\\
- & \frac{\epsilon_m}{(2l-1+m)!}\sum_{l'=1}^{\infty} \sideset{}{'}\sum_{m'=0}^{2l'-1}
\frac{\widetilde{U}_{2l-1,m;2l'-1,m'}}{(2l'-1-m')!} \bar{B}_{2l'-1,m'}
,
\end{split}
\end{equation}
where $\widetilde{U}_{2l-1,m;2l'-1,m'} = (\bar{U}_{2(l+l'-1)}^{|m'-m|}+\bar{U}_{2(l+l'-1)}^{m+m'})/2 $ and the prime at the summation indicates that only $m'$ divisible by $4$ are included. On the other hand, Eq.~\eqref{eq:n-final} gives
\begin{equation}
\label{eq:C_in_D}
\begin{split}
C_{2l-1,m}& = \frac{\epsilon_m \sqrt{4l-1} }{\bar{\kappa}} \frac{\sqrt{(2l-1-m)!}}{\sqrt{(2l-1+m)!}} 
\\
\times \sum_{l'=1}^{\infty} \sum_{m'=0}^{2l'-1} &\frac{\sqrt{(2l'-1+m')!}}{\sqrt{(2l'-1-m')!}}  \frac{\widetilde{G}_{2l-1,m;2l'-1,m'}}{ \sqrt{4 l' -1}}  D_{2l'-1,m'},
\end{split}
\end{equation}
where $\widetilde{G}_{2l-1,m;2l'-1,m'}=(\bar{G}_{2l-1,m;2l'-1,m'}+ \bar{G}_{2l-1,m;2l'-1,-m'})/2$.

By inserting Eqs.~\eqref{eq:A_in_B} and~\eqref{eq:C_in_D} into the first equation of~\eqref{eq:BC-matched}, we obtain following linear equations to an arbitrary order in the angular momentum $\langle 2 l-1,m \rangle$ sector:
\begin{widetext}
\begin{equation}
\label{eq:B-D-1}
\begin{split}
-H \delta_{l1}\delta_{m0}=& \sum_{l'=1}^{\infty} \sum_{m'=0}^{2l'-1} \left[ \delta_{l l'}  \delta_{m m'} \frac{2l}{\epsilon_m a^{2l+1} } +\frac{(2l-1) a^{2l} \widetilde{U}_{2l-1,m;2l'-1,m'}}{(2l-1+m)! (2l'-1-m')!} \right] \bar{B}_{2l'-1,m'} 
\\
+&  \sum_{l'=1}^{\infty} \sum_{m'=0}^{2l'-1} \left[ \delta_{l l'}  \delta_{m m'} \frac{\bar{\kappa} n'_l(\bar{\kappa} \bar{a})}{\epsilon_m}  + \frac{\sqrt{4l-1} }{\sqrt{4l'-1} }  \frac{\sqrt{(2l-1-m)!}}{\sqrt{(2l-1+m)!}}  \frac{\sqrt{(2l'-1+m')!}}{\sqrt{(2l'-1-m')!}} \widetilde{G}_{2l-1,m;2l'-1,m'} j'_l(\bar{\kappa} \bar{a}) \right] D_{2l'-1,m'}. 
\end{split}
\end{equation}
Similarly,  the second equation in~\eqref{eq:BC-matched} becomes:
\begin{equation}
\label{eq:B-D-2}
\begin{split}
-H \delta_{l1}\delta_{m0}=& l (2l-1) \sum_{l'=1}^{\infty} \sum_{m'=0}^{2l'-1} \left[ \frac{a^{2l} \widetilde{U}_{2l-1,m;2l'-1,m'}}{(2l-1+m)! (2l'-1-m')!} - \delta_{l l'}  \delta_{m m'} \frac{1}{\epsilon_m a^{2l+1} }  \right] \bar{B}_{2l'-1,m'} 
\\
+&  \sum_{l'=1}^{\infty} \sum_{m'=0}^{2l'-1} \left[ \delta_{l l'}  \delta_{m m'} \frac{\bar{\kappa} }{\epsilon_m} \left( \frac{n'_l(\bar{\kappa} \bar{a})}{\xi} - l (2l-1) \frac{n_l(\bar{\kappa} \bar{a})}{\bar{\kappa} \bar{a}} \right) \right.
\\
& \qquad \qquad \left.+ \frac{\sqrt{4l-1} }{\sqrt{4l'-1} }  \frac{\sqrt{(2l-1-m)!}}{\sqrt{(2l-1+m)!}}  \frac{\sqrt{(2l'-1+m')!}}{\sqrt{(2l'-1-m')!}} \widetilde{G}_{2l-1,m;2l'-1,m'} \left( \frac{j'_l(\bar{\kappa} \bar{a})}{\xi} - l (2l-1) \frac{j_l(\bar{\kappa} \bar{a})}{\bar{\kappa} \bar{a}}  \right) \right] D_{2l'-1,m'}. 
\end{split}
\end{equation}
\end{widetext}

We have reduced four sets of linear equations to two sets. For arbitrary order of the angular momentum cutoff, $L$ and $M$, given by $L =\max(2l-1) = \max(2l'-1)$ and $M =\max(2m-1) = \max(2m'-1)$, by truncating the higher angular momentum terms, it is straightforward to set up a square matrix associated with these two sets of linear equations, once the values of the structure factors $\widetilde{U}_{2l-1,m;2l'-1,m'}$ and $\widetilde{G}_{2l-1,m;2l'-1,m'}$ are evaluated. After that, inverting the matrix gives us the value of $B_{1,0}^{\langle L,M \rangle}$ and, hence, the complex conductivity from Eq.~\eqref{eq:conductivity_rescaled}.

The dielectric constant can then be obtained from the complex conductivity with the following relation:
\begin{equation}
\sigma_s (\omega)= \sigma_s'(\omega) + i\sigma_s''(\omega)
\equiv \varepsilon_0 \omega \epsilon_s''(\omega)- i \varepsilon_0 \omega \epsilon_s'(\omega),
\end{equation}
where $\sigma_s'$ and $\sigma_s''$ are real and imaginary part of the conductivity, while $\epsilon_s'$ and $\epsilon_s''$ are real and imaginary part of the dielectric constant. We then have $\epsilon_s'=-\sigma_s''/(\varepsilon_0\omega)$. It is convenient and also useful to subtract the DC contribution from the imaginary part of dielectric constant and define the parameter $\tilde{\epsilon}_s''$ as
\begin{equation}
\tilde{\epsilon}_s''=\frac{\sigma_s'(\omega)- \sigma_s'(\omega=0)}{\varepsilon_0 \omega}.
\end{equation}
The peak position of $\tilde{\epsilon}_s''(\omega)$ as a function of frequency directly reflects the characteristic relaxation frequency~\cite{Chew1982}.

Our focus here is how the sphere size $a$ and the lattice spacing $b$ affect the dielectric response. Hence, we will not discuss the effect of the surface charge strength $\Omega$ and the ionic concentration $N_0$ of the electrolyte. Qualitatively, their effects follow trends similar to the single particle response discussed in Ref.~\cite{Chew1982} and~\cite{Hou2017}. A larger amount of surface charge $\Omega$ will lead to a stronger response, while higher ionic concentration suppresses the dielectric enhancement. Because the filling fraction is given by $f= (4\pi/3) \bar{a}^3$, varying $f$ while keeping $a$ (or $b$) fixed is equivalent to varying $b$ (or $a$). Throughout our discussion, we will use the following values for the parameters: the diffusion coefficient is given by $D=1.334\times 10^{-9}$ m$^2$/s; the relative permittivity of the electrolyte is given by $\epsilon'_w=80$; the surface charge density is given by $\Omega = 0.5$ nm$^{-2}$; the intrinsic ion density is given by $N_0=1.03 \times 10^{25}$ m$^{-3}$; and the electrolyte conductivity is given by $\sigma'_w=0.2$ S/m. Here, the values of the ion density and the electrolyte conductivity roughly correspond to those of an electrolyte with $1$ part per thousand in weight of NaCl.  

\begin{figure}
	\includegraphics[width=0.45\textwidth]{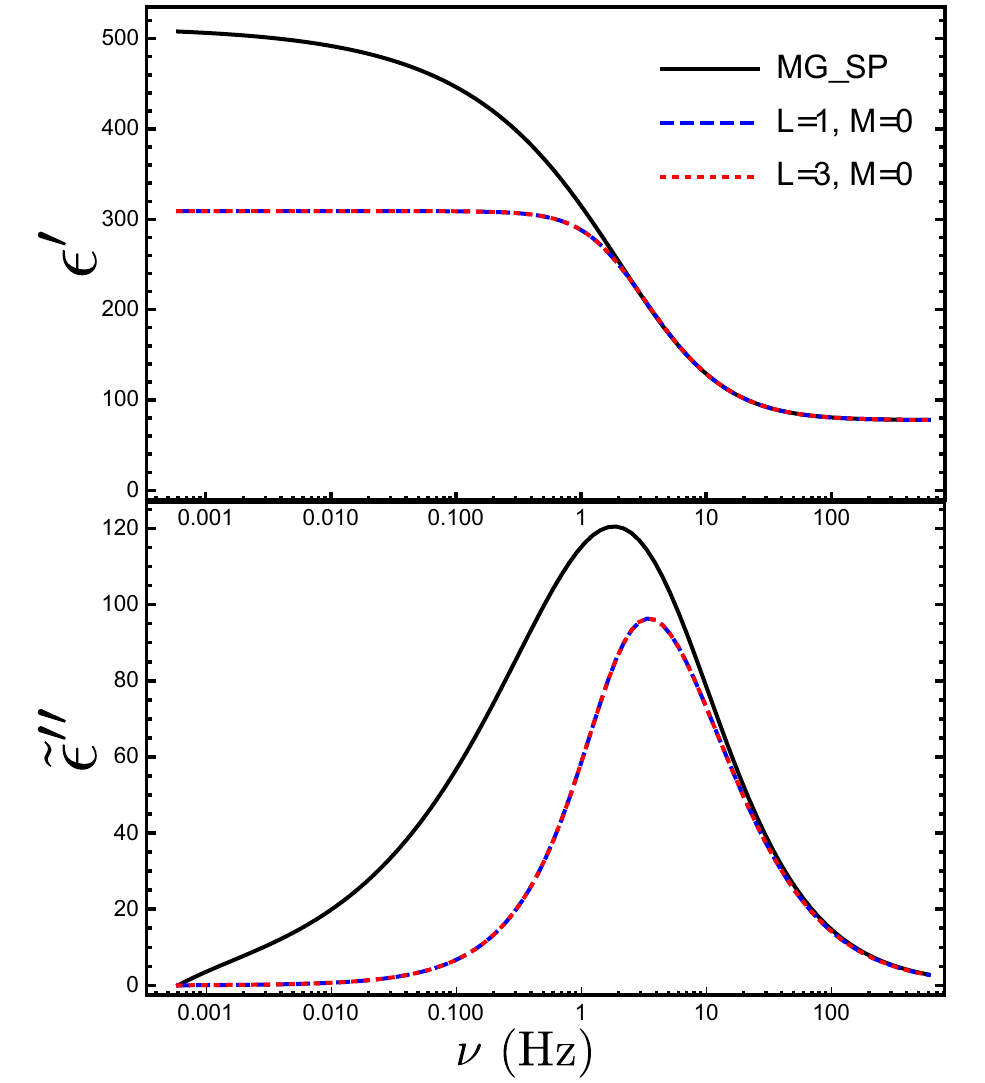}
	\caption{The dispersion of $\epsilon'$ and $\tilde{\epsilon}''$ with different angular momentum cutoffs $\langle L,M \rangle$ is shown as a function of frequency. The black solid curve shows the response using the Maxwell Garnett mixing formula with the single-particle polarization coefficient. The size of the spheres is $a=10$ $\mu$m and the volume fraction of the charged spheres is $f=2$ \%.} 
	\label{fig:Eps_convergence_1}
\end{figure}

\begin{figure}
	\includegraphics[width=0.45\textwidth]{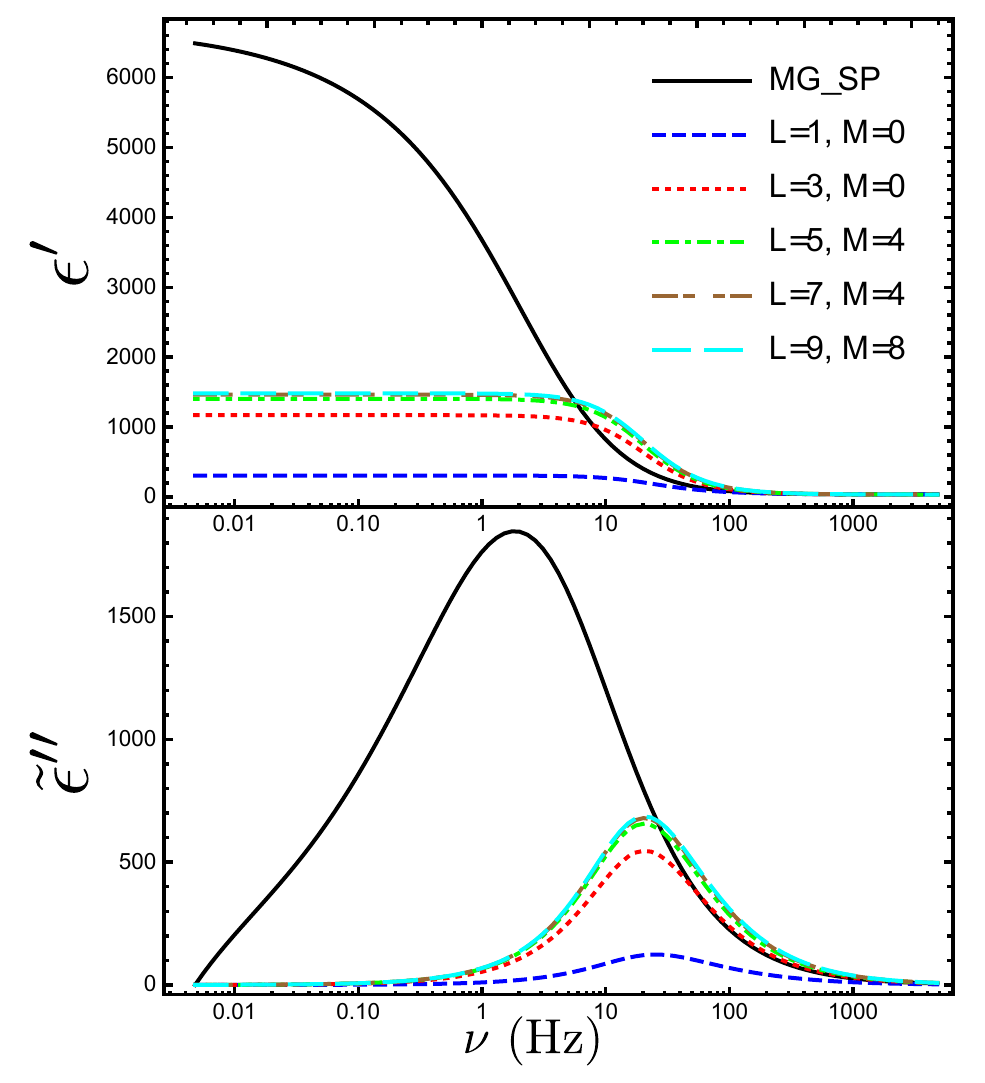}
	\caption{The dispersions of $\epsilon'$ and $\tilde{\epsilon}''$ with different angular momentum cutoffs $\langle L,M \rangle$ is shown as a function of frequency. The black solid curve shows the response using the Maxwell Garnett mixing formula with the single-particle polarization coefficient. The size of the spheres is $a=10$ $\mu$m and the volume fraction of the charged spheres is $f=45$ \%.} 
	\label{fig:Eps_convergence_2}
\end{figure}

We are now in a position to evaluate ${\epsilon'_s}^{\langle L,M\rangle}(\omega)$ and $\tilde{\epsilon}''_{s}{}^{\langle L,M\rangle}(\omega)$  from the $B_{1,0}^{\langle L,M\rangle}$ evaluated to an arbitrary cutof,f $\langle L,M\rangle$, in the angular momentum. Let us first investigate the convergence behavior of ${\epsilon'_s}^{\langle L,M\rangle}(\omega)$ and $\tilde{\epsilon}''_{s}{}^{\langle L,M\rangle}(\omega)$, beginning with lower order and then considering higher order values of the cutoff f $\langle L,M\rangle$. In Figs.~\ref{fig:Eps_convergence_1} and~\ref{fig:Eps_convergence_2}, we show the dispersion of $\epsilon'$ and $\tilde{\epsilon}''$ as a function of frequency for different angular momentum cutoffs, up to $\langle L,M \rangle = \langle 9,8 \rangle $. The size of the spheres is taken to be $a=10$ $\mu$m. The volume fractions of the charged spheres are $f=2$~\% and $f=45$~\% for Figs.~\ref{fig:Eps_convergence_1} and~\ref{fig:Eps_convergence_2}, respectively. For smaller volume fractions of charged spheres, c.f. Fig.~\ref{fig:Eps_convergence_1}, the dielectric response calculated from only the lowest order of the angular momentum terms already gives the convergent result because ${\epsilon'_s}^{\langle 1,0 \rangle}(\omega)$  and $\tilde{\epsilon}''_{s}{}^{\langle 1,0 \rangle}(\omega)$ are virtually the same as ${\epsilon'_s}^{\langle 3,0 \rangle}(\omega)$ and $\tilde{\epsilon}''_{s}{}^{\langle 3,0 \rangle}(\omega)$. For higher volume fractions of charged spheres, we need to include higher angular momentum corrections, as shown in Fig.~\ref{fig:Eps_convergence_2}. However, it converges rapidly because ${\epsilon'_s}^{\langle 3,0\rangle}(\omega)$ and $\tilde{\epsilon}''_{s}{}^{\langle 3,0\rangle}(\omega)$ are very similar to ${\epsilon'_s}^{\langle 9,8\rangle}(\omega)$ and $\tilde{\epsilon}''_{s}{}^{\langle 9,8\rangle}(\omega)$. In general, the convergence becomes slower with higher volume fractions of charged spheres, i.e., it requires higher angular momentum corrections to reach the convergent result. In this work, we will restrict our discussion to cases with $f< 45$ \%, where the angular momentum cutoff to order $\langle L,M \rangle= \langle 9, 8\rangle$  already gives convergent results. Our experience shows that higher order corrections are needed when $f> 45$ \%.

We also show, in both Figs.~\ref{fig:Eps_convergence_1} and~\ref{fig:Eps_convergence_2}, the comparison between our results and those from the Maxwell Garnett mixing formula, depicted by the black solid curve labeled "MG\_SP", given in Eq.~\eqref{eq:M-G-conductivity}. We found that the dielectric response of charged spheres arranged in a cubic lattice will reduce to the Maxwell Garnett mixing formula in two limits: (1) the high frequency regime and (2) lower volume fractions of charged spheres. This is consistent with the discussion in the previous section. In general, we observe the  trend that the characteristic relaxation frequency for charged spheres arranged in a cubic lattice is higher than that predicted by the Maxwell Garnett mixing formula. This shows that the presence of nearby charged spheres strongly affects the neutral current flow that is induced by a single charged sphere.  It is an indication that the polarization response of a single particle given in Eq.~\eqref{eq:polarization_coefficient-single_particle} may not represent the essential physics for more densely packed systems. 

\begin{figure}
	\includegraphics[width=0.45\textwidth]{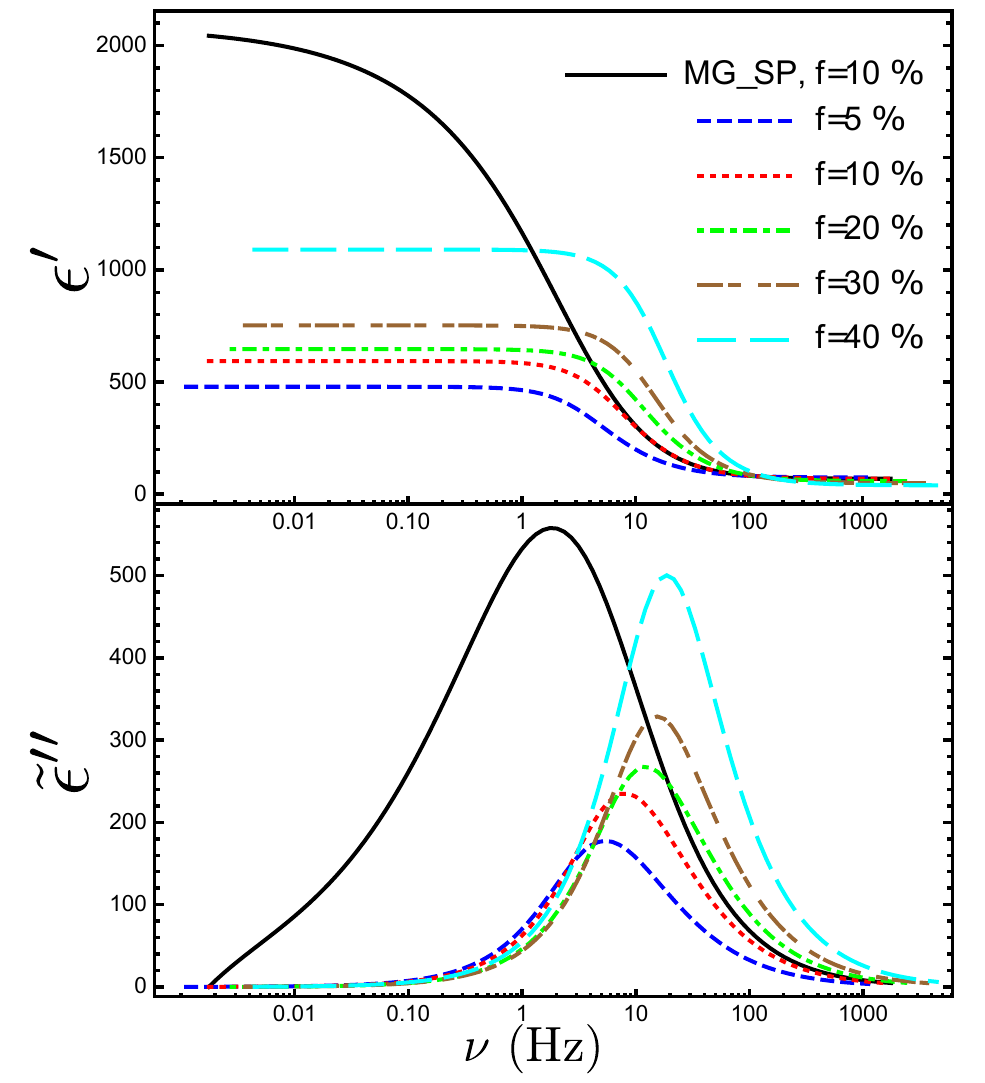}
	\caption{The dispersion of $\epsilon'$ and $\tilde{\epsilon}''$ for different volume fractions, $f$, of charged spheres is shown as a function of frequency. The size of the spheres is $a=10$ $\mu$m. The angular momentum cutoff is $\langle L,M \rangle= \langle 9, 8\rangle$. For comparison, the black solid curve shows the prediction from the Maxwell Garnett mixing formula in Eq.~\eqref{eq:M-G-conductivity} with $f=10$ \%.} 
	\label{fig:Eps_f_vary}
\end{figure}

In Fig.~\ref{fig:Eps_f_vary}, we illustrate how the dielectric response of charged spheres arranged in a simple cubic lattice varies with the change in volume fraction. Again, the size of the spheres is $a=10$ $\mu$m. We choose the angular momentum cutoff to be $\langle L,M \rangle= \langle 9, 8\rangle$, which has  reasonable convergence, as shown in Fig.~\ref{fig:Eps_convergence_2}. For comparison, we also show the predicted dielectric response for the corresponding Maxwell Garnett formula with $f=10$ \%. It is important to note that the characteristic frequency does not depend on the volume fraction of the charged spheres in the Maxwell Garnett formula. Only the height of the peak and the strenght of the dielectric enhancement will depend on the volume fraction~\cite{Chew1982,Hou2017}. With increasing volume fractions of charged spheres, the characteristic frequencies predicted by our model  (the position of the peak of the $\tilde{\epsilon}''$) shift toward higher values and, hence, away from those given by the Maxwell Garnett formula. This is a clear indication that inter-particle interactions can and will make the dielectric response deviate strongly from the single-particle physics. 

\begin{figure}
	\includegraphics[width=0.45\textwidth]{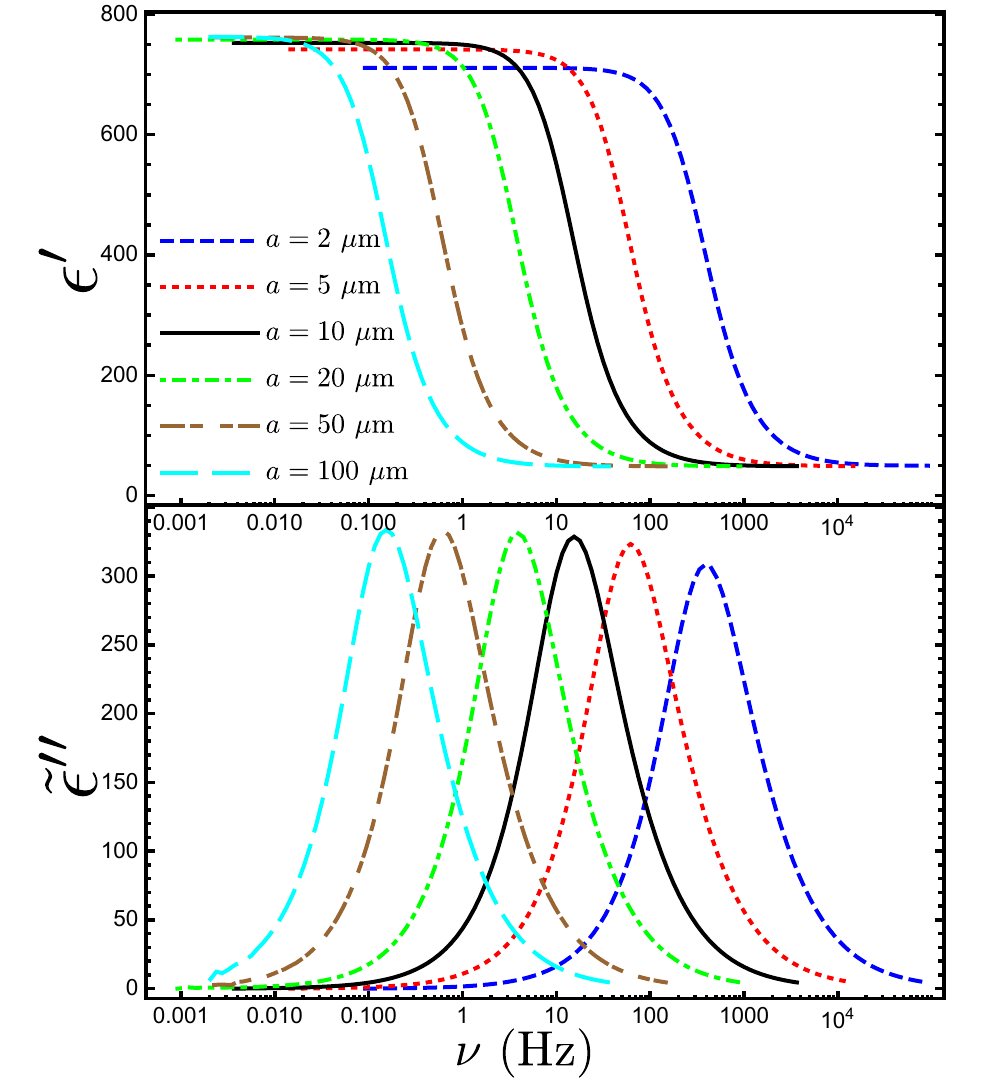}
	\caption{The dispersion of $\epsilon'$ and $\tilde{\epsilon}''$ for charged spheres with different radii, $2$, $5$, $10$, $20$, $50$, $100$ $\mu$m, is shown as a function of frequency. The volume fraction of the spheres is fixed at $f=30$ \%. The angular momentum cutoff is $\langle L,M \rangle= \langle 9, 8\rangle$.} 
	\label{fig:Eps_f30_a_vary}
\end{figure}

Finally, we show in Fig.~\ref{fig:Eps_f30_a_vary} the size dependence of the dielectric response for a fixed volume fraction of charged spheres. As expected, the characteristic frequency decreases with increasing size of the spheres, which is qualitatively consistent with the prediction from the Maxwell Garnett formula with the single-particle polarization coefficient. However,  the resulting characteristic frequencies in our model are roughly one order of magnitude higher than that predicted by the Maxwell Garnett formula as shown in fig.~\ref{fig:Eps_f_vary}.

\section{Discussion}
\label{sec:discussion}

The anomalous low-frequency dielectric response of charged particles immersed in an electrolyte has been observed in various types of systems, from  suspensions of polystyrene latex particles,  emulsions of fat particles~\cite{Schwan1962}, and suspensions of clay particles~\cite{LEROY2017,Hou2018} to complicated rock formations~\cite{Wait1959}. Often, the explication of this phenomenon begins with the single-particle polarization response, similar to that given in Eq.~\eqref{eq:polarization_coefficient-single_particle}. Then, a representative model of the system is established using various types of effective medium approximations~\cite{Chew1982,Hou2017,Chassagne2008,Stroud1978,Zhdanov2008} which aim to capture the influence of the particles' interactions on the dielectric response. This approach is particularly vulnerable to the specific mechanism discussed in this work and first proposed in Ref.~\cite{Chew1982} because the non-localized neutral current in the electrolyte is responsible for the low-frequency dielectric enhancement.

As alluded to in Sec.~\ref{sec:model}, the non-localized neutral current extends farther away from the charged sphere as the frequency is decreased. Hence, the presence of other charged spheres could affect the neutral current and influence the dielectric response. As a result, our solution for the dielectric response of the system of interest, charged spheres arranged in a simple cubic lattice and immersed in an electrolyte, serves as an important step in properly modeling  more densely packed suspensions or even jams. Because the method illustrated in this paper can be carried out to arbitrary order in the angular momentum cutoff with proper numerical implementation, the predicted dielectric response from the theory effectively becomes exact once  proper convergence is observed, c.f. Figs.~\ref{fig:Eps_convergence_1} and~\ref{fig:Eps_convergence_2}. Even though our modeled system is rather restrictive given the strict periodicity, the fact that an exact solution can be obtained eliminates  ambiguities originating from the effective medium approximation, and it allows us to draw valuable insights.  

As illustrated in Figs.~\ref{fig:Eps_convergence_1}--\ref{fig:Eps_f_vary}, the dielectric response of our model exhibits qualitatively different behavior from that in which only the single-particle physics is taken into account. First and foremost, the tendency to induce a strong dielectric enhancement competes with a suppression effect due to the presence of other spheres. We thus observe a weaker low-frequency dielectric enhancement according to our solution compared to that from the Maxwell Garnett formula. In addition, because the neutral current extends much farther at lower frequencies, the suppression effect becomes stronger in that regime, with the characteristic relaxation frequency shifting to a higher value compared to that predicted by the single-particle polarization response. In general,  larger volume fractions of charged spheres lead to a higher characteristic relaxation frequency in our model, c.f. Fig.~\ref{fig:Eps_f_vary}. Combined with the observation in Fig.~\ref{fig:Eps_f30_a_vary} that the characteristic frequency also depends on the size of the charged sphere given a fixed volume fraction, we can conclude that both the sphere size and the separation distance between spheres will influence the resulting characteristic relaxation frequency.

The introduction of the second length scale is not unexpected. For a single particle, the low frequency dielectric response is dominated by the neutral current that flows in response to the density gradient of the ions that built up on the surface of the sphere~\cite{Chew1982}. Because the positive and negative charges collect at the opposite ends of the sphere, the size of the sphere is the natural length scale. Once there is more than one sphere, the neutral current can also flow in response to the density gradient of the ions that have collected on the surfaces of adjacent spheres. This gives a second length scale associated with the distance between the two spheres. 

It is worthwhile pointing out that the Maxwell Garnett formula with the single-particle polarization response does qualitatively describe the measured dielectric response of spherical polystyrene particles, effectively charged spheres, suspended in a KCl electrolyte~\cite{Schwan1962,Chew1982}. Even though our model allows us to take into account the inter-particle effect, the predicted dielectric response from our solution becomes less consistent with the measured dielectric response found in Ref.~\cite{Schwan1962}.
 
Two possible scenarios can lead to the discrepancy between our theory and the experiment. First, the packing of the polystyrene spheres in the suspension is likely to be more random than our simple cubic array. Namely, there can be some lattice effects in our calculations that lead to a qualitative difference from more randomized systems. However, unlike in the case of electron localization versus conduction in periodic crystals where the periodic scatterers introduce interference effects that are qualitatively different from those of random obstacles, the ``wave vector'', $\kappa$ (below Eq.~\eqref{eq:green-def}) in our case is complex with the real and imaginary parts having the same amplitude. As a result, the effective wavelength of the oscillation is the same as the length scale of the exponential decay. Thus, we do not expect a qualitative and sizable difference between the responses of periodic and randomly packed structures due to interference. Indeed, the consideration of different lattice structures, such as BCC or FCC structures, could help us to identify the lattice effects. Given our argument that the non-localized neutral current should be more suppressed at lower frequencies, we still expect that the dielectric enhancement will, in general, be suppressed, with the characteristic relaxation frequency shifted to higher values compared to the single-particle response. Second, the mechanism discussed in this paper may not truly capture the physics responsible for the dielectric enhancement observed in Ref.~\cite{Schwan1962}. This implies that the qualitative agreement between the single particle response and the experiment might simply be a coincidence~\cite{Chew1982}, and other types of mechanisms, such as those discussed in Ref.~\cite{Schwarz1962}, are responsible for the observed phenomenon. Additional quantitative experiments are required to understand the mechanism behind the observed dielectric enhancement.

\section{Summary}
\label{sec:summary}

In summary, we have presented a formalism for solving for the complex dielectric constant 
of a system consisting of charged spheres arranged in a simple cubic lattice 
and immersed in an electrolyte. The influence of the periodicity on the 
dielectric response is recast into two periodic conditions for two of the variables: 
the electric potential obeying Laplace's equation, and the ion concentration 
described by  Helmholtz's equation. For Laplace's equation, we employed 
the method developed by Lord Rayleigh for computing the complex conductivity 
ofs system with periodically arranged dielectric spheres embedded in 
another dielectric material~\cite{Rayleigh1892}. We then applied the KKR 
method developed for band structure calculations to derive the periodic 
condition for the Helmholtz equation~\cite{Korringa,KohnRostoker}. Two other 
sets of boundary conditions are given at the outer surface of the electric 
double-layer as derived by Fixman for the single-particle polarization response 
of a charged sphere, which effectively encodes all the information about the 
interaction between the electric potential and the ion density. Combining all these 
efforts, we established a numerical scheme for computing the dispersion of the complex 
conductivity to arbitrary order in the angular momentum cutoff.  

Our solution shows that the Maxwell Garnett formula, together with the single-particle polarization coefficient, does not properly describe the dielectric 
response of our system even when only the lowest order $\langle L=1,M=0 \rangle$ 
corrections are taken into account, c.f. Fig.~\ref{fig:Eps_f_vary}, even for 
volume fractions of the spheres  as low as two percent.  With  increasing volume 
fractions of charged spheres, we found that higher angular momentum corrections 
are required to obtain  convergence to the exact solution, as indicated in 
Fig.~\ref{fig:Eps_convergence_2}. Also, as illustrated in 
Fig.~\ref{fig:Eps_f_vary}, the dielectric response of the modeled system will 
become qualitatively more distinct from the single-particle response when the
charged spheres occupy a higher volume fraction. Because our modeled system 
allows us to rigorously account for the influence of multiple particle on the 
dielectric responses, our findings indicate that the presence of nearby 
charged spheres will strongly modify the polarization response given by a single,
stand alone, charged sphere. Hence, a densely packed system will give rise to 
a dielectric response that is qualitatively and quantitatively different from a 
dilute suspension. Further investigations on periodic structures, such as BCC 
and FCC lattice, will further our understanding of the dielectric response of densely packed systems.

\section*{Acknowledgments}

The authors would like to thank P. N. Sen for pointing out the omission of the topic of our studies in the literature and for valuable discussions. The authors also thank N. Seleznev and the Petrophysics program at Schlumberger-Doll Research for the support of this work.

\appendix

\section{Evaluation of structure factors}
\label{app:structure-factor}

To complete the periodic boundary conditions discussed in Sec.~\ref{sec:Periodic_BCs}, we need to evaluate two sets of structure factors: $U_l^m$ and $G_{l,m;l',m'}$. The former can be directly evaluated using the definition in Eq.~\eqref{eq:phi-structure}. Below, we will provide some evaluated values of $U_{l}^{m}$, which have been documented in Ref.~\onlinecite{McPhedran1} with less precision. The evaluation of $G_{l,m;l',m'}$ is more involved. The direct use of Eq.~\eqref{eq:structure-factors} is not feasible as it will converge very slowly. In addition, because $G_{l,m;l',m'}$ is a function of $\kappa \equiv \sqrt{i \omega/D}$, it requires computing a larger number of them to construct a faithful dielectric dispersion. Fortunately, an efficient method for computing the structure factor has been developed by Ham and Segall~\cite{Ham1961}. We will review their formulae for computing the structure factors.

\subsection{Evaluation of $U_l^m$}

In principle, the structure factor $U_l^m$ can be directly evaluated by Eq.~\eqref{eq:phi-structure}, with one important twist. Although the shape of the considered system does not affect the evaluated value for most of the $U_l^m$ as long as a large enough number of terms are included in the summation, only the value of $U_2^0$ depends on the choice of shape. One can demonstrate that a system with full spherical or cubic symmetry will have $U_2^0=0$. However, as pointed out by Lord Rayleigh and later argued by McPhedran and McKenzie~\cite{Rayleigh1892,McPhedran1}, the long needle is the representative shape. In this case, the value of $U_2^0$ is simply $4\pi/3$. (Note the slight difference between our definition of $U_l^m$ and theirs.) 

Here, we list some values of $U_{l}^{m}$ evaluated with Eq.~\eqref{eq:phi-structure}:
\begin{center}
\begin{tabular}{|c||c||c|} 
	\hline
	\scriptsize $U_{2}^{0}= 4\pi/3 $ & &\\ 
	\hline 
	\scriptsize $U_{4}^{0}=74.5956 $ & & \\ 
	\hline
	\scriptsize $U_{6}^{0}=412.797 $ & \scriptsize $U_{6}^{4}=-2889.58$ &\\
	\hline
	\scriptsize $U_{8}^{0}=131414.7 $ & \scriptsize $U_{8}^{4}=131414.7$ & \\
	\hline 
	\scriptsize $U_{10}^{0}=3.662272\times10^6 $ & \scriptsize $U_{10}^{4}=-8.056998\times10^6$ & \scriptsize $U_{10}^{8}= -1.36969\times10^8$ \\
	\hline
	\scriptsize $U_{12}^{0}= 1.384915\times10^9$ &  \scriptsize $U_{12}^{4}= 8.189911\times 10^8$ &  \scriptsize $U_{12}^{8}= 8.176006\times10^9$ \\
    \hline
\end{tabular} 
\end{center}
The terms with higher values of $l$ and $m$ can be easily computed by summing a few terms close to the origin in Eq.~\eqref{eq:phi-structure}.

\subsection{Evaluation of $G_{l,m;l',m'}$}

As noted by Kohn and Rostoker, the evaluation of $G_{l,m;l',m'}$ with Eq.~\eqref{eq:structure-factors} converges very slowly~\cite{KohnRostoker}. Hence, an alternative formula has been developed for effectively computing the structure factor. In the most general case, $G_{l,m;l',m'}$ depends both on the energy $\kappa$ and the Bloch wave vector $\vec{k}$. Because we have $\vec{k}=0$ due to the strictly periodic condition in our problem and because we are interested in the simple cubic lattice, we will only review formulas that fit our purposes.

Following the observation of Kohn and Rostoker~\cite{KohnRostoker}, the structure factor $\bar{G}_{l,m;l',m'}$ can be expressed as a finite summation of another form of structure factor, $\bar{\mathcal{D}}_{\mathcal{L},\mathcal{M}}$, as follows:
\begin{equation}
\bar{G}_{l,m;l',m'} = 4\pi i^{(l-l')} \sum_{\mathcal{L}} \frac{1}{i^{\mathcal{L}}} \bar{\mathcal{D}}_{\mathcal{L},m-m'} \mathcal{C}_{\mathcal{L},m-m';l,m;l',m'},
\end{equation} 
where the sum of $\mathcal{L}$ runs over the values $|l-l'|$, $|l-l'|+2$, $\cdots$ $|l+l'|$, and
\begin{equation}
 \mathcal{C}_{\mathcal{L},\mathcal{M};l,m;l',m'}= \int d\Omega Y_{\mathcal{L}\mathcal{M}} (\theta,\phi) Y_{lm}^* (\theta,\phi)Y_{l'm'} (\theta,\phi).
\end{equation}
It is useful to note that $\mathcal{C}_{\mathcal{L},\mathcal{M};l,m;l',m'}=0$ when $\mathcal{M}\neq m-m'$ due to angular momentum conservation. Again, the bar over the symbols $\bar{G}_{l,m;l',m'}$ and $\bar{\mathcal{D}}_{\mathcal{L},\mathcal{M}}$ indicates that variables are rescaled by the lattice spacing $b$ and is dimensionless.

As shown by Ham and Segall~\cite{Ham1961}, the structure factors $\bar{\mathcal{D}}_{\mathcal{L},\mathcal{M}}$ can be casted into three sets of infinite summations as follows:
\begin{equation}
\bar{\mathcal{D}}_{\mathcal{L},\mathcal{M}}= \bar{\mathcal{D}}_{\mathcal{L},\mathcal{M}}^{(1)}+\bar{\mathcal{D}}_{\mathcal{L},\mathcal{M}}^{(2)}+\bar{\mathcal{D}}_{\mathcal{L},\mathcal{M}}^{(3)},
\end{equation}
where the speed of convergence for each term is controlled by an arbitrary dimensionless parameter $\bar{\eta}$. Remarkably, even though values of $\bar{\mathcal{D}}_{\mathcal{L},\mathcal{M}}^{(\alpha)}$ with $\alpha=1,2,3$ depend on $\bar{\eta}$, adding them to obtain $\bar{\mathcal{D}}_{\mathcal{L},\mathcal{M}}$ yields a result that is independent from the choice of $\bar{\eta}$.

The first summation is given by 
\begin{equation}
\bar{\mathcal{D}}_{\mathcal{L},\mathcal{M}}^{(1)}= -4 \pi \frac{i^{\mathcal{L}} }{\bar{\kappa}^{\mathcal{L}} }  e^{\bar{\kappa}^2/\bar{\eta} }  \sum_{n} \frac{ |\vec{\bar{K}}_n|^{\mathcal{L}}  e^{-|\vec{\bar{K}}_n|^{2} / \bar{\eta} } } {|\vec{\bar{K}}_n|^{2} - \bar{\kappa}^2 } Y_{\mathcal{L}\mathcal{M}}^* (\theta_n,\phi_n), 
\end{equation}
where $\vec{\bar{K}}_n=2\pi \left( n_x \hat{x} + n_y \hat{y} + n_z \hat{z}\right)$ are rescaled reciprocal lattice vectors, and $\theta_n$ and $\phi_n$ are their corresponding polar angles. Here, $n_{x,y,z}$ runs over all integers. The second summation reads as
\begin{equation}
\begin{split}
 \bar{\mathcal{D}}_{\mathcal{L},\mathcal{M}}^{(2)}=& -\frac{2^{\mathcal{L} +1} }{\sqrt{\pi} \bar{\kappa}^{\mathcal{L}} } \sum_{s\neq 0} |\vec{\bar{r}}_s|^{\mathcal{L}} Y_{\mathcal{L}\mathcal{M}}^* (\theta_s,\phi_s)
 \\
 &\qquad \qquad \quad \times \int_{\sqrt{\bar{\eta}}/2}^{\infty} d\xi \xi^{2 \mathcal{L}}
 e^{- \xi^{2} |\vec{\bar{r}}_s|^{2} + \kappa^2/(4 )\xi^{2} }  , 
 \end{split}
\end{equation}
where $\vec{\bar{r}}_s = s_x \hat{x} + s_y \hat{y} + s_z \hat{z}$ are the rescaled lattice vectors, and $\theta_s$ and $\phi_s$ are their corresponding polar angles. In the summation, $s_{x,y,z}$ runs over all integers except $s_x=s_y=s_z=0$ as indicated by $s\neq 0$. The third summation is given by
\begin{equation}
\bar{\mathcal{D}}_{\mathcal{L},\mathcal{M}}^{(3)}= - \delta_{\mathcal{L} 0}  \delta_{\mathcal{M} 0}  \frac{\sqrt{\bar{\eta}}}{2 \pi}  \sum_{\alpha=0}^{\infty} \frac{\left(\kappa^2/\bar{\eta}\right)^\alpha }{(2\alpha-1) \alpha !} , 
\end{equation}
which only contributes to the structure factor when $\mathcal{L}=\mathcal{M}=0$. Finally, it is worthwhile to mention that $\bar{\mathcal{D}}_{\mathcal{L},\mathcal{M}}^{(1)}$ will converge faster (requiring less terms away from the origin of the reciprocal lattice) with a smaller value of $\bar{\eta}$ while $\bar{\mathcal{D}}_{\mathcal{L},\mathcal{M}}^{(2)}$ will converge faster (requiring less terms away from the origin of the lattice) with a larger value of $\bar{\eta}$. Numerically, an effective way for choosing a proper value of $\bar{\eta}$ is thus very important for the faster overall convergence.

\bibliographystyle{apsrev4-1}
\bibliography{biblio}

\begin{thebibliography}{44}%
\makeatletter
\providecommand \@ifxundefined [1]{%
 \@ifx{#1\undefined}
}%
\providecommand \@ifnum [1]{%
 \ifnum #1\expandafter \@firstoftwo
 \else \expandafter \@secondoftwo
 \fi
}%
\providecommand \@ifx [1]{%
 \ifx #1\expandafter \@firstoftwo
 \else \expandafter \@secondoftwo
 \fi
}%
\providecommand \natexlab [1]{#1}%
\providecommand \enquote  [1]{``#1''}%
\providecommand \bibnamefont  [1]{#1}%
\providecommand \bibfnamefont [1]{#1}%
\providecommand \citenamefont [1]{#1}%
\providecommand \href@noop [0]{\@secondoftwo}%
\providecommand \href [0]{\begingroup \@sanitize@url \@href}%
\providecommand \@href[1]{\@@startlink{#1}\@@href}%
\providecommand \@@href[1]{\endgroup#1\@@endlink}%
\providecommand \@sanitize@url [0]{\catcode `\\12\catcode `\$12\catcode
  `\&12\catcode `\#12\catcode `\^12\catcode `\_12\catcode `\%12\relax}%
\providecommand \@@startlink[1]{}%
\providecommand \@@endlink[0]{}%
\providecommand \url  [0]{\begingroup\@sanitize@url \@url }%
\providecommand \@url [1]{\endgroup\@href {#1}{\urlprefix }}%
\providecommand \urlprefix  [0]{URL }%
\providecommand \Eprint [0]{\href }%
\providecommand \doibase [0]{http://dx.doi.org/}%
\providecommand \selectlanguage [0]{\@gobble}%
\providecommand \bibinfo  [0]{\@secondoftwo}%
\providecommand \bibfield  [0]{\@secondoftwo}%
\providecommand \translation [1]{[#1]}%
\providecommand \BibitemOpen [0]{}%
\providecommand \bibitemStop [0]{}%
\providecommand \bibitemNoStop [0]{.\EOS\space}%
\providecommand \EOS [0]{\spacefactor3000\relax}%
\providecommand \BibitemShut  [1]{\csname bibitem#1\endcsname}%
\let\auto@bib@innerbib\@empty
\bibitem [{\citenamefont {Schwan}\ \emph {et~al.}(1962)\citenamefont {Schwan},
  \citenamefont {Schwarz}, \citenamefont {Maczuk},\ and\ \citenamefont
  {Pauly}}]{Schwan1962}%
  \BibitemOpen
  \bibfield  {author} {\bibinfo {author} {\bibfnamefont {H.~P.}\ \bibnamefont
  {Schwan}}, \bibinfo {author} {\bibfnamefont {G.}~\bibnamefont {Schwarz}},
  \bibinfo {author} {\bibfnamefont {J.}~\bibnamefont {Maczuk}}, \ and\ \bibinfo
  {author} {\bibfnamefont {H.}~\bibnamefont {Pauly}},\ }\href@noop {}
  {\bibfield  {journal} {\bibinfo  {journal} {The Journal of Physical
  Chemistry}\ }\textbf {\bibinfo {volume} {66}},\ \bibinfo {pages} {2626}
  (\bibinfo {year} {1962})}\BibitemShut {NoStop}%
\bibitem [{\citenamefont {Chassagne}\ \emph {et~al.}(2003)\citenamefont
  {Chassagne}, \citenamefont {Bedeaux}, \citenamefont {Van~der Ploeg},\ and\
  \citenamefont {Koper}}]{Chassagne2003}%
  \BibitemOpen
  \bibfield  {author} {\bibinfo {author} {\bibfnamefont {C.}~\bibnamefont
  {Chassagne}}, \bibinfo {author} {\bibfnamefont {D.}~\bibnamefont {Bedeaux}},
  \bibinfo {author} {\bibfnamefont {J.~P.~M.}\ \bibnamefont {Van~der Ploeg}}, \
  and\ \bibinfo {author} {\bibfnamefont {G.~J.~M.}\ \bibnamefont {Koper}},\
  }\href@noop {} {\bibfield  {journal} {\bibinfo  {journal} {Langmuir}\
  }\textbf {\bibinfo {volume} {19}},\ \bibinfo {pages} {3619} (\bibinfo {year}
  {2003})}\BibitemShut {NoStop}%
\bibitem [{\citenamefont {Schwarz}(1962)}]{Schwarz1962}%
  \BibitemOpen
  \bibfield  {author} {\bibinfo {author} {\bibfnamefont {G.}~\bibnamefont
  {Schwarz}},\ }\href@noop {} {\bibfield  {journal} {\bibinfo  {journal} {The
  Journal of Physical Chemistry}\ }\textbf {\bibinfo {volume} {66}},\ \bibinfo
  {pages} {2636} (\bibinfo {year} {1962})}\BibitemShut {NoStop}%
\bibitem [{\citenamefont {Dukhin}\ and\ \citenamefont
  {Shilov}(1974)}]{Dukhin1974}%
  \BibitemOpen
  \bibfield  {author} {\bibinfo {author} {\bibfnamefont {S.}~\bibnamefont
  {Dukhin}}\ and\ \bibinfo {author} {\bibfnamefont {V.}~\bibnamefont
  {Shilov}},\ }\href@noop {} {\emph {\bibinfo {title} {Dielectric phenomena and
  Double layer in Disperse system and Polyelectrolytes}}}\ (\bibinfo
  {publisher} {Wiley, New York},\ \bibinfo {year} {1974})\BibitemShut {NoStop}%
\bibitem [{\citenamefont {Fixman}(1980)}]{Fixman}%
  \BibitemOpen
  \bibfield  {author} {\bibinfo {author} {\bibfnamefont {M.}~\bibnamefont
  {Fixman}},\ }\href@noop {} {\bibfield  {journal} {\bibinfo  {journal} {The
  Journal of Chemical Physics}\ }\textbf {\bibinfo {volume} {72}},\ \bibinfo
  {pages} {5177} (\bibinfo {year} {1980})}\BibitemShut {NoStop}%
\bibitem [{\citenamefont {Dukhin}\ and\ \citenamefont
  {Shilov}(1980)}]{Dukhin1980}%
  \BibitemOpen
  \bibfield  {author} {\bibinfo {author} {\bibfnamefont {S.}~\bibnamefont
  {Dukhin}}\ and\ \bibinfo {author} {\bibfnamefont {V.}~\bibnamefont
  {Shilov}},\ }\href@noop {} {\bibfield  {journal} {\bibinfo  {journal}
  {Advances in Colloid and Interface Science}\ }\textbf {\bibinfo {volume}
  {13}},\ \bibinfo {pages} {153 } (\bibinfo {year} {1980})}\BibitemShut
  {NoStop}%
\bibitem [{\citenamefont {DeLacey}\ and\ \citenamefont
  {White}(1981)}]{DeLacey1981}%
  \BibitemOpen
  \bibfield  {author} {\bibinfo {author} {\bibfnamefont {E.~H.~B.}\
  \bibnamefont {DeLacey}}\ and\ \bibinfo {author} {\bibfnamefont {L.~R.}\
  \bibnamefont {White}},\ }\href@noop {} {\bibfield  {journal} {\bibinfo
  {journal} {J. Chem. Soc.{,} Faraday Trans. 2}\ }\textbf {\bibinfo {volume}
  {77}},\ \bibinfo {pages} {2007} (\bibinfo {year} {1981})}\BibitemShut
  {NoStop}%
\bibitem [{\citenamefont {Chew}\ and\ \citenamefont
  {Sen}(1982{\natexlab{a}})}]{Chew1982}%
  \BibitemOpen
  \bibfield  {author} {\bibinfo {author} {\bibfnamefont {W.~C.}\ \bibnamefont
  {Chew}}\ and\ \bibinfo {author} {\bibfnamefont {P.~N.}\ \bibnamefont {Sen}},\
  }\href@noop {} {\bibfield  {journal} {\bibinfo  {journal} {The Journal of
  Chemical Physics}\ }\textbf {\bibinfo {volume} {77}},\ \bibinfo {pages}
  {4683} (\bibinfo {year} {1982}{\natexlab{a}})}\BibitemShut {NoStop}%
\bibitem [{\citenamefont {Chassagne}\ and\ \citenamefont
  {Bedeaux}(2008)}]{Chassagne2008}%
  \BibitemOpen
  \bibfield  {author} {\bibinfo {author} {\bibfnamefont {C.}~\bibnamefont
  {Chassagne}}\ and\ \bibinfo {author} {\bibfnamefont {D.}~\bibnamefont
  {Bedeaux}},\ }\href@noop {} {\bibfield  {journal} {\bibinfo  {journal}
  {Journal of Colloid and Interface Science}\ }\textbf {\bibinfo {volume}
  {326}},\ \bibinfo {pages} {240 } (\bibinfo {year} {2008})}\BibitemShut
  {NoStop}%
\bibitem [{\citenamefont {Chassagne}(2013)}]{Chassagne2013}%
  \BibitemOpen
  \bibfield  {author} {\bibinfo {author} {\bibfnamefont {C.}~\bibnamefont
  {Chassagne}},\ }\href@noop {} {\bibfield  {journal} {\bibinfo  {journal}
  {International Journal of Thermophysics}\ }\textbf {\bibinfo {volume} {34}},\
  \bibinfo {pages} {1239} (\bibinfo {year} {2013})}\BibitemShut {NoStop}%
\bibitem [{\citenamefont {Hou}\ \emph {et~al.}(2017)\citenamefont {Hou},
  \citenamefont {Freed},\ and\ \citenamefont {Sen}}]{Hou2017}%
  \BibitemOpen
  \bibfield  {author} {\bibinfo {author} {\bibfnamefont {C.-Y.}\ \bibnamefont
  {Hou}}, \bibinfo {author} {\bibfnamefont {D.~E.}\ \bibnamefont {Freed}}, \
  and\ \bibinfo {author} {\bibfnamefont {P.~N.}\ \bibnamefont {Sen}},\
  }\href@noop {} {\bibfield  {journal} {\bibinfo  {journal} {Phys. Rev. E}\
  }\textbf {\bibinfo {volume} {95}},\ \bibinfo {pages} {042601} (\bibinfo
  {year} {2017})}\BibitemShut {NoStop}%
\bibitem [{\citenamefont {Hou}\ \emph {et~al.}(2018)\citenamefont {Hou},
  \citenamefont {Feng}, \citenamefont {Seleznev},\ and\ \citenamefont
  {Freed}}]{Hou2018}%
  \BibitemOpen
  \bibfield  {author} {\bibinfo {author} {\bibfnamefont {C.-Y.}\ \bibnamefont
  {Hou}}, \bibinfo {author} {\bibfnamefont {L.}~\bibnamefont {Feng}}, \bibinfo
  {author} {\bibfnamefont {N.}~\bibnamefont {Seleznev}}, \ and\ \bibinfo
  {author} {\bibfnamefont {D.~E.}\ \bibnamefont {Freed}},\ }\href@noop {}
  {\bibfield  {journal} {\bibinfo  {journal} {Journal of Colloid and Interface
  Science}\ }\textbf {\bibinfo {volume} {525}},\ \bibinfo {pages} {62 }
  (\bibinfo {year} {2018})}\BibitemShut {NoStop}%
\bibitem [{\citenamefont {Leroy}\ \emph {et~al.}(2017)\citenamefont {Leroy},
  \citenamefont {Weigand}, \citenamefont {Mériguet}, \citenamefont
  {Zimmermann}, \citenamefont {Tournassat}, \citenamefont {Fagerlund},
  \citenamefont {Kemna},\ and\ \citenamefont {Huisman}}]{LEROY2017}%
  \BibitemOpen
  \bibfield  {author} {\bibinfo {author} {\bibfnamefont {P.}~\bibnamefont
  {Leroy}}, \bibinfo {author} {\bibfnamefont {M.}~\bibnamefont {Weigand}},
  \bibinfo {author} {\bibfnamefont {G.}~\bibnamefont {Mériguet}}, \bibinfo
  {author} {\bibfnamefont {E.}~\bibnamefont {Zimmermann}}, \bibinfo {author}
  {\bibfnamefont {C.}~\bibnamefont {Tournassat}}, \bibinfo {author}
  {\bibfnamefont {F.}~\bibnamefont {Fagerlund}}, \bibinfo {author}
  {\bibfnamefont {A.}~\bibnamefont {Kemna}}, \ and\ \bibinfo {author}
  {\bibfnamefont {J.~A.}\ \bibnamefont {Huisman}},\ }\href@noop {} {\bibfield
  {journal} {\bibinfo  {journal} {Journal of Colloid and Interface Science}\
  }\textbf {\bibinfo {volume} {505}},\ \bibinfo {pages} {1093 } (\bibinfo
  {year} {2017})}\BibitemShut {NoStop}%
\bibitem [{\citenamefont {Sihvola}(1999)}]{Sihvola1999}%
  \BibitemOpen
  \bibfield  {author} {\bibinfo {author} {\bibfnamefont {A.}~\bibnamefont
  {Sihvola}},\ }\href@noop {} {\emph {\bibinfo {title} {Electromagnetic mixing
  formulas and applications}}}\ (\bibinfo  {publisher} {The institute of
  Electric Engineers, London, UK},\ \bibinfo {year} {1999})\BibitemShut
  {NoStop}%
\bibitem [{\citenamefont {Jackson}(1998)}]{Jackson1998}%
  \BibitemOpen
  \bibfield  {author} {\bibinfo {author} {\bibfnamefont {J.~D.}\ \bibnamefont
  {Jackson}},\ }\href@noop {} {\emph {\bibinfo {title} {Classical
  Electrodynamics, Third Edition}}}\ (\bibinfo  {publisher} {John Wiley and
  Sons, Inc., New York},\ \bibinfo {year} {1998})\BibitemShut {NoStop}%
\bibitem [{\citenamefont {Landau}\ and\ \citenamefont
  {Lifshitz}(1984)}]{Landau1984}%
  \BibitemOpen
  \bibfield  {author} {\bibinfo {author} {\bibfnamefont {L.~D.}\ \bibnamefont
  {Landau}}\ and\ \bibinfo {author} {\bibfnamefont {E.~M.}\ \bibnamefont
  {Lifshitz}},\ }\href@noop {} {\emph {\bibinfo {title} {Electrodynamics of
  Continuous Media, Second Edition (Course of Theoretical Physics): Volume
  8}}}\ (\bibinfo  {publisher} {Butterworth-Heinemann},\ \bibinfo {year}
  {1984})\BibitemShut {NoStop}%
\bibitem [{\citenamefont {Chew}\ and\ \citenamefont
  {Sen}(1982{\natexlab{b}})}]{Chew1982a}%
  \BibitemOpen
  \bibfield  {author} {\bibinfo {author} {\bibfnamefont {W.~C.}\ \bibnamefont
  {Chew}}\ and\ \bibinfo {author} {\bibfnamefont {P.~N.}\ \bibnamefont {Sen}},\
  }\href@noop {} {\bibfield  {journal} {\bibinfo  {journal} {The Journal of
  Chemical Physics}\ }\textbf {\bibinfo {volume} {77}},\ \bibinfo {pages}
  {2042} (\bibinfo {year} {1982}{\natexlab{b}})}\BibitemShut {NoStop}%
\bibitem [{\citenamefont {Garnett}(1904)}]{Garnett1904}%
  \BibitemOpen
  \bibfield  {author} {\bibinfo {author} {\bibfnamefont {J.~C.~M.}\
  \bibnamefont {Garnett}},\ }\href@noop {} {\bibfield  {journal} {\bibinfo
  {journal} {Philosophical Transactions of the Royal Society of London A:
  Mathematical, Physical and Engineering Sciences}\ }\textbf {\bibinfo {volume}
  {203}},\ \bibinfo {pages} {385} (\bibinfo {year} {1904})}\BibitemShut
  {NoStop}%
\bibitem [{\citenamefont {Garnett}(1906)}]{Garnett1906}%
  \BibitemOpen
  \bibfield  {author} {\bibinfo {author} {\bibfnamefont {J.~C.~M.}\
  \bibnamefont {Garnett}},\ }\href@noop {} {\bibfield  {journal} {\bibinfo
  {journal} {Philosophical Transactions of the Royal Society of London A:
  Mathematical, Physical and Engineering Sciences}\ }\textbf {\bibinfo {volume}
  {205}},\ \bibinfo {pages} {237} (\bibinfo {year} {1906})}\BibitemShut
  {NoStop}%
\bibitem [{\citenamefont {Bruggeman}(1935)}]{Bruggeman1935}%
  \BibitemOpen
  \bibfield  {author} {\bibinfo {author} {\bibfnamefont {D.~A.~G.}\
  \bibnamefont {Bruggeman}},\ }\href@noop {} {\bibfield  {journal} {\bibinfo
  {journal} {Annalen der Physik}\ }\textbf {\bibinfo {volume} {416}},\ \bibinfo
  {pages} {636} (\bibinfo {year} {1935})}\BibitemShut {NoStop}%
\bibitem [{\citenamefont {Milton}(1964)}]{Milton2002}%
  \BibitemOpen
  \bibfield  {author} {\bibinfo {author} {\bibfnamefont {G.~W.}\ \bibnamefont
  {Milton}},\ }\href@noop {} {\emph {\bibinfo {title} {The Theory of
  Composites}}}\ (\bibinfo  {publisher} {Cambridge University Press, Cambridge,
  UK},\ \bibinfo {year} {1964})\BibitemShut {NoStop}%
\bibitem [{\citenamefont {Sen}\ and\ \citenamefont {Kan}(1987)}]{SenKan1}%
  \BibitemOpen
  \bibfield  {author} {\bibinfo {author} {\bibfnamefont {P.~N.}\ \bibnamefont
  {Sen}}\ and\ \bibinfo {author} {\bibfnamefont {R.}~\bibnamefont {Kan}},\
  }\href@noop {} {\bibfield  {journal} {\bibinfo  {journal} {Physical Review
  Letters}\ }\textbf {\bibinfo {volume} {58}},\ \bibinfo {pages} {778}
  (\bibinfo {year} {1987})}\BibitemShut {NoStop}%
\bibitem [{\citenamefont {Kan}\ and\ \citenamefont {Sen}(1987)}]{SenKan2}%
  \BibitemOpen
  \bibfield  {author} {\bibinfo {author} {\bibfnamefont {R.}~\bibnamefont
  {Kan}}\ and\ \bibinfo {author} {\bibfnamefont {P.~N.}\ \bibnamefont {Sen}},\
  }\href@noop {} {\bibfield  {journal} {\bibinfo  {journal} {The Journal of
  Chemical Physics}\ }\textbf {\bibinfo {volume} {86}},\ \bibinfo {pages}
  {5748} (\bibinfo {year} {1987})}\BibitemShut {NoStop}%
\bibitem [{\citenamefont {Sen}(1987)}]{Sen1987}%
  \BibitemOpen
  \bibfield  {author} {\bibinfo {author} {\bibfnamefont {P.~N.}\ \bibnamefont
  {Sen}},\ }\href@noop {} {\bibfield  {journal} {\bibinfo  {journal} {The
  Journal of Chemical Physics}\ }\textbf {\bibinfo {volume} {87}},\ \bibinfo
  {pages} {4100} (\bibinfo {year} {1987})}\BibitemShut {NoStop}%
\bibitem [{\citenamefont {Gu}\ and\ \citenamefont {Yu}(1991)}]{Gu1991}%
  \BibitemOpen
  \bibfield  {author} {\bibinfo {author} {\bibfnamefont {G.}~\bibnamefont
  {Gu}}\ and\ \bibinfo {author} {\bibfnamefont {K.~W.}\ \bibnamefont {Yu}},\
  }\href@noop {} {\bibfield  {journal} {\bibinfo  {journal} {Journal of Applied
  Physics}\ }\textbf {\bibinfo {volume} {70}},\ \bibinfo {pages} {4476}
  (\bibinfo {year} {1991})}\BibitemShut {NoStop}%
\bibitem [{\citenamefont {Ni}\ \emph {et~al.}(2002)\citenamefont {Ni},
  \citenamefont {Gu},\ and\ \citenamefont {Chen}}]{Ni2002}%
  \BibitemOpen
  \bibfield  {author} {\bibinfo {author} {\bibfnamefont {F.-S.}\ \bibnamefont
  {Ni}}, \bibinfo {author} {\bibfnamefont {G.-Q.}\ \bibnamefont {Gu}}, \ and\
  \bibinfo {author} {\bibfnamefont {K.-M.}\ \bibnamefont {Chen}},\ }\href@noop
  {} {\bibfield  {journal} {\bibinfo  {journal} {Chinese Physics Letters}\
  }\textbf {\bibinfo {volume} {19}},\ \bibinfo {pages} {1550} (\bibinfo {year}
  {2002})}\BibitemShut {NoStop}%
\bibitem [{\citenamefont {Rayleigh}(1892)}]{Rayleigh1892}%
  \BibitemOpen
  \bibfield  {author} {\bibinfo {author} {\bibfnamefont {L.}~\bibnamefont
  {Rayleigh}},\ }\href@noop {} {\bibfield  {journal} {\bibinfo  {journal} {The
  London, Edinburgh, and Dublin Philosophical Magazine and Journal of Science}\
  }\textbf {\bibinfo {volume} {34}},\ \bibinfo {pages} {481} (\bibinfo {year}
  {1892})}\BibitemShut {NoStop}%
\bibitem [{\citenamefont {McPhedran}\ and\ \citenamefont
  {McKenzie}(1978)}]{McPhedran1}%
  \BibitemOpen
  \bibfield  {author} {\bibinfo {author} {\bibfnamefont {R.}~\bibnamefont
  {McPhedran}}\ and\ \bibinfo {author} {\bibfnamefont {D.}~\bibnamefont
  {McKenzie}},\ }\href@noop {} {\bibfield  {journal} {\bibinfo  {journal}
  {Proceedings of the Royal Society of London A: Mathematical, Physical and
  Engineering Sciences}\ }\textbf {\bibinfo {volume} {359}},\ \bibinfo {pages}
  {45} (\bibinfo {year} {1978})}\BibitemShut {NoStop}%
\bibitem [{\citenamefont {McKenzie}\ \emph {et~al.}(1978)\citenamefont
  {McKenzie}, \citenamefont {McPhedran},\ and\ \citenamefont
  {Derrick}}]{McPhedran2}%
  \BibitemOpen
  \bibfield  {author} {\bibinfo {author} {\bibfnamefont {D.}~\bibnamefont
  {McKenzie}}, \bibinfo {author} {\bibfnamefont {R.}~\bibnamefont {McPhedran}},
  \ and\ \bibinfo {author} {\bibfnamefont {G.}~\bibnamefont {Derrick}},\
  }\href@noop {} {\bibfield  {journal} {\bibinfo  {journal} {Proceedings of the
  Royal Society of London A: Mathematical, Physical and Engineering Sciences}\
  }\textbf {\bibinfo {volume} {362}},\ \bibinfo {pages} {211} (\bibinfo {year}
  {1978})}\BibitemShut {NoStop}%
\bibitem [{\citenamefont {Korringa}(1947)}]{Korringa}%
  \BibitemOpen
  \bibfield  {author} {\bibinfo {author} {\bibfnamefont {J.}~\bibnamefont
  {Korringa}},\ }\href@noop {} {\bibfield  {journal} {\bibinfo  {journal}
  {Physica}\ }\textbf {\bibinfo {volume} {13}},\ \bibinfo {pages} {392}
  (\bibinfo {year} {1947})}\BibitemShut {NoStop}%
\bibitem [{\citenamefont {Kohn}\ and\ \citenamefont
  {Rostoker}(1954)}]{KohnRostoker}%
  \BibitemOpen
  \bibfield  {author} {\bibinfo {author} {\bibfnamefont {W.}~\bibnamefont
  {Kohn}}\ and\ \bibinfo {author} {\bibfnamefont {N.}~\bibnamefont
  {Rostoker}},\ }\href@noop {} {\bibfield  {journal} {\bibinfo  {journal}
  {Physical Review}\ }\textbf {\bibinfo {volume} {94}},\ \bibinfo {pages}
  {1111} (\bibinfo {year} {1954})}\BibitemShut {NoStop}%
\bibitem [{\citenamefont {Qian}\ and\ \citenamefont {Sen}(2015)}]{jiang}%
  \BibitemOpen
  \bibfield  {author} {\bibinfo {author} {\bibfnamefont {J.}~\bibnamefont
  {Qian}}\ and\ \bibinfo {author} {\bibfnamefont {P.~N.}\ \bibnamefont {Sen}},\
  }\href@noop {} {\enquote {\bibinfo {title} {Universal dielectric enhancement
  from externally induced double layer without $\zeta$-potential},}\ }
  (\bibinfo {year} {2015}),\ \Eprint {http://arxiv.org/abs/1510.06724}
  {arxiv:1510.06724 [physics.class-ph]} \BibitemShut {NoStop}%
\bibitem [{\citenamefont {Fixman}(1983)}]{Fixman1983}%
  \BibitemOpen
  \bibfield  {author} {\bibinfo {author} {\bibfnamefont {M.}~\bibnamefont
  {Fixman}},\ }\href@noop {} {\bibfield  {journal} {\bibinfo  {journal} {The
  Journal of Chemical Physics}\ }\textbf {\bibinfo {volume} {78}},\ \bibinfo
  {pages} {1483} (\bibinfo {year} {1983})}\BibitemShut {NoStop}%
\bibitem [{Note1()}]{Note1}%
  \BibitemOpen
  \bibinfo {note} {The condition is, as always, that the wavelength of the
  electromagnetic perturbation, which for a MHz drive is hundreds of meters, is
  much larger than the system's geometric features.}\BibitemShut {Stop}%
\bibitem [{Note2()}]{Note2}%
  \BibitemOpen
  \bibinfo {note} {The proof is standard and involves integrating the vector
  field around a loop, a rectangle that is long in the tangential direction and
  very short, across $\delta $ in the radial direction, and that the short
  radial line integrals can be neglected. However, it may be objected that
  within the double layer there is a huge gradient in the potential because of
  the build-up of charges within the short distance $\delta $, so the radial
  integrals may be non-negligible. However, from Eq.~\protect \textup {\hbox
  {\mathsurround \z@ \protect \normalfont (\ignorespaces \ref
  {eq:perturbation}\unskip \@@italiccorr )}} we can see that the radial
  gradient of the \protect \emph {chemical potential} $\mu $ is actually of the
  order of the perturbation, due to the absence of currents in equilibrium.
  Thus the usual argument indeed works, as long as the right order of limits is
  taken.}\BibitemShut {Stop}%
\bibitem [{\citenamefont {Bourg}\ and\ \citenamefont
  {Sposito}(2011)}]{Bourg2011}%
  \BibitemOpen
  \bibfield  {author} {\bibinfo {author} {\bibfnamefont {I.~C.}\ \bibnamefont
  {Bourg}}\ and\ \bibinfo {author} {\bibfnamefont {G.}~\bibnamefont
  {Sposito}},\ }\href@noop {} {\bibfield  {journal} {\bibinfo  {journal}
  {Journal of Colloid and Interface Science}\ }\textbf {\bibinfo {volume}
  {360}},\ \bibinfo {pages} {701 } (\bibinfo {year} {2011})}\BibitemShut
  {NoStop}%
\bibitem [{\citenamefont {Morse}\ and\ \citenamefont
  {Feshbach}(1964)}]{Morse1953}%
  \BibitemOpen
  \bibfield  {author} {\bibinfo {author} {\bibfnamefont {P.~M.}\ \bibnamefont
  {Morse}}\ and\ \bibinfo {author} {\bibfnamefont {H.}~\bibnamefont
  {Feshbach}},\ }\href@noop {} {\emph {\bibinfo {title} {Methods of theoretical
  physics: Part II, page 1292}}}\ (\bibinfo  {publisher} {McGraw-Hill Book
  Company, INC.},\ \bibinfo {year} {1964})\BibitemShut {NoStop}%
\bibitem [{\citenamefont {Landau}\ and\ \citenamefont
  {Lifshitz}(1965)}]{Landau1965}%
  \BibitemOpen
  \bibfield  {author} {\bibinfo {author} {\bibfnamefont {L.~D.}\ \bibnamefont
  {Landau}}\ and\ \bibinfo {author} {\bibfnamefont {E.~M.}\ \bibnamefont
  {Lifshitz}},\ }\href@noop {} {\emph {\bibinfo {title} {Quantum Mechanics,
  Second Edition (Course of Theoretical Physics): Volume 3}}}\ (\bibinfo
  {publisher} {Butterworth-Heinemann},\ \bibinfo {year} {1965})\BibitemShut
  {NoStop}%
\bibitem [{Note3()}]{Note3}%
  \BibitemOpen
  \bibinfo {note} {The derivation here is not entirely rigorous. The Green's
  function $\protect \mathcal {G}_{E,\protect \mathaccentV
  {vec}17E{k}}(\protect \mathaccentV {vec}17E{r},\protect \mathaccentV
  {vec}17E{r}\protect \tmspace +\thinmuskip {.1667em}')$ becomes singular as
  $\protect \mathaccentV {vec}17E{r}$ and $\protect \mathaccentV
  {vec}17E{r}\protect \tmspace +\thinmuskip {.1667em}'$ coincide, so the
  integrals in Eqs.~\protect \textup {\hbox {\mathsurround \z@ \protect
  \normalfont (\ignorespaces \ref {eq:Green-derive1}\unskip \@@italiccorr )}}
  and~\protect \textup {\hbox {\mathsurround \z@ \protect \normalfont
  (\ignorespaces \ref {eq:Green-derive2}\unskip \@@italiccorr )}} require more
  care, i.e., by restricting the range of integration to near, but not at, the
  surfaces of the spheres, c.f.~\cite {KohnRostoker} Appendix I for more
  details}\BibitemShut {NoStop}%
\bibitem [{\citenamefont {Davis}(1971)}]{Davis1971}%
  \BibitemOpen
  \bibfield  {author} {\bibinfo {author} {\bibfnamefont {H.~L.}\ \bibnamefont
  {Davis}},\ }\href@noop {} {\emph {\bibinfo {title} {Computational Methods in
  Band Theory (Edited by Marcus P.M., Janak J.F. and Williams A.R.): Efficient
  Numerical Techniques for the calculation of KKR structure constant}}}\
  (\bibinfo  {publisher} {Springer, Boston, MA},\ \bibinfo {year}
  {1971})\BibitemShut {NoStop}%
\bibitem [{\citenamefont {Ham}\ and\ \citenamefont {Segall}(1961)}]{Ham1961}%
  \BibitemOpen
  \bibfield  {author} {\bibinfo {author} {\bibfnamefont {F.~S.}\ \bibnamefont
  {Ham}}\ and\ \bibinfo {author} {\bibfnamefont {B.}~\bibnamefont {Segall}},\
  }\href@noop {} {\bibfield  {journal} {\bibinfo  {journal} {Phys. Rev.}\
  }\textbf {\bibinfo {volume} {124}},\ \bibinfo {pages} {1786} (\bibinfo {year}
  {1961})}\BibitemShut {NoStop}%
\bibitem [{\citenamefont {Wait}(1959)}]{Wait1959}%
  \BibitemOpen
  \bibfield  {author} {\bibinfo {author} {\bibfnamefont {J.~R.}\ \bibnamefont
  {Wait}},\ }\href@noop {} {\emph {\bibinfo {title} {Overvoltage Research and
  Geophysical Applications}}}\ (\bibinfo  {publisher} {Pergamon Press},\
  \bibinfo {year} {1959})\BibitemShut {NoStop}%
\bibitem [{\citenamefont {Stroud}(1975)}]{Stroud1978}%
  \BibitemOpen
  \bibfield  {author} {\bibinfo {author} {\bibfnamefont {D.}~\bibnamefont
  {Stroud}},\ }\href@noop {} {\bibfield  {journal} {\bibinfo  {journal} {Phys.
  Rev. B}\ }\textbf {\bibinfo {volume} {12}},\ \bibinfo {pages} {3368}
  (\bibinfo {year} {1975})}\BibitemShut {NoStop}%
\bibitem [{\citenamefont {Zhdanov}(2008)}]{Zhdanov2008}%
  \BibitemOpen
  \bibfield  {author} {\bibinfo {author} {\bibfnamefont {M.}~\bibnamefont
  {Zhdanov}},\ }\href@noop {} {\bibfield  {journal} {\bibinfo  {journal}
  {GEOPHYSICS}\ }\textbf {\bibinfo {volume} {73}},\ \bibinfo {pages} {F197}
  (\bibinfo {year} {2008})}\BibitemShut {NoStop}%
\end{thebibliography}%
\end{document}